\documentclass[]{raa}    
\usepackage{graphicx,times}             
\usepackage{natbib}
\usepackage{amssymb,amsmath}
\usepackage{grffile}

\bibpunct{(}{)}{;}{a}{}{,}

\usepackage[a4paper=true,pagebackref=true]{hyperref}
\hypersetup{colorlinks = true, linkcolor = blue, anchorcolor = red, citecolor = blue, filecolor = red, pagecolor = red, urlcolor = green}

\newcommand{\K}{\ensuremath{\, {\rm K}}}

\begin{document}

\title{The Electromagnetic Characteristics of the Tianlai Cylindrical Pathfinder Array}

 \volnopage{Vol.0 (20xx) No.0, 000--000}      
 \setcounter{page}{1}          

\author{Shijie Sun \inst{1,2}, Jixia Li \inst{1}, 
Fengquan Wu \inst{1},
Peter Timbie \inst{3},
Reza Ansari \inst{4}, 
Jingchao Geng \inst{5},
Huli Shi \inst{1},
Albert Stebbins \inst{6},
Yougang Wang \inst{1},
Juyong Zhang \inst{7},
Xuelei Chen \inst{1,2,8,9}
}
  \institute{National Astronomical Observatories, Chinese Academy of Sciences, Beijing 100101, China; \\
    \and
    University of Chinese Academy of Sciences, Beijing 100049, China\\
    \and  
    Department of Physics, University of Wisconsin Madison, 1150 University Ave, Madison WI 53703, USA\\
    \and
    IJC Lab, CNRS/IN2P3 \& Universit\'e Paris-Saclay, 15 rue Georges Cl\'emenceau, 91405 Orsay, France\\
    \and  
    The 54th Research Institute, China Electronics Technology Group Corporation, Shijiazhuang, Hebei 050051, China\\
    \and
    Fermi National Accelerator Laboratory, P.O. Box 500, Batavia IL 60510-5011, USA\\
    \and
    School of Mechanical Engineering, Hangzhou Dianzi University, Hangzhou 310017, China\\
    \and 
    Department of Physics, College of Sciences, Northeastern University, Shenyang 110819, China\\
    \and
    Center of High Energy Physics, Peking University, Beijing 100871, China\\
    {\it Emails: xuelei@cosmology.bao.ac.cn; sshj@nao.cas.cn}\\
\vs\no
   {\small Received~~20xx month day; accepted~~20xx~~month day}
}



\abstract{A great challenge for 21 cm intensity mapping experiments is the strong foreground radiation which is orders of magnitude brighter than the 21cm signal. Removal of the foreground takes advantage of the fact that its frequency spectrum is smooth while the redshifted 21cm signal spectrum is stochastic. However, a complication is the non-smoothness of the instrument response. This paper describes the electromagnetic simulation of the Tianlai cylinder array, a pathfinder for 21 cm intensity mapping experiments. Due to the vast scales involved, a direct simulation requires large amount of computing resources. We have made the simulation practical by using a combination of methods: first simulate a single feed, then an array of feed units, finally with the feed array and a cylindrical reflector together, to obtain the response for a single cylinder. We studied its radiation pattern, bandpass response and the effects of mutual coupling between feed units, and compared the results with observation. Many features seen in the measurement result are well reproduced in the simulation, especially the oscillatory features which are associated with the standing waves on the reflector. The mutual coupling between feed units is quantified with S-parameters, which decrease as the distance between the two feeds increases. Based on the simulated S-parameters, we estimate the correlated noise which has been seen in the visibility data, the results show very good agreement with the data in both magnitude and frequency structures. These results provide useful insights on the problem of 21cm signal extraction for real instruments.  
}	

   \authorrunning{Shijie Sun et al.}            
   \titlerunning{Electromagnetic Characteristics of Tianlai Cylindrical Array}  

   \maketitle

\section{Introduction}

The 21 cm line of neutral hydrogen (HI) provides a unique tool for probing the Universe throughout most of its history, including the dark ages, cosmic dawn, 
the epoch of reionization (EoR), and the post-reionization Universe. In the last decade, a number of experimental efforts have been dedicated to 21 cm 
tomography observations. These includes the EoR experiments, such as LOFAR \citep{LOFAR2013}, MWA \citep{MWA2013}, PAPER \citep{PAPER2010}, HERA \citep{HERA2017}), and SKA-low, as well as post-reionization large scale structure experiments, which aim to probe the dark energy by measuring the baryon acoustic oscillation (BAO) features in the power spectrum, such as CHIME (The Canadian Hydrogen Intensity Mapping Experiment) \citep{amiri2022detection,amiri2022overview,vanderlinde2014canadian}, HIRAX (The Hydrogen Intensity and Real-time Analysis eXperiment) \citep{crichton2018hydrogen,newburgh2016hirax}, BINGO (The Baryon Acoustic Oscillations from Integrated Neutral Gas Observations) \citep{abdalla2021bingo}, MeerKAT \citep{jonas2016meerkat}, ASKAP (Australian Square Kilometre Array Pathfinder) \citep{hotan2021australian, johnston2008science}, FAST \citep{Hu:2019okh} and SKA-mid \citep{Koopmans2015,SKA:2018ckk}. Future experiments of even greater scale, such as PUMA (Packed Ultra-wideband Mapping Array) \citep{bandura2019packed,oconnor2020baryon}, are also being discussed. 

The Tianlai project is also a 21 cm experiment designed to survey the large scale structure by intensity mapping of the redshifted 21 cm line \citep{chen2012tianlai,chen2015tianlai,chen2015AAPPS,Xu2015,2016RAA....16..158Z,Das2018}. To resolve the BAO peaks, compact interferometer arrays with longest baseline of at least a hundred meter are required \citep{chang2008baryon,ansari2008reconstruction,seo2010ground}. As a first step, the current pathfinder experiment will test the basic principles and key technologies of the 21 cm intensity mapping method. The Tianlai experiment has two pathfinder arrays, both located at the radio-quiet Hongliuxia site in Xinjiang, west China. The cylinder pathfinder arrays presently consist of three 15 m$\times$40 m cylindrical reflectors with no moving parts \citep{li2020tianlai}, and the dish  array consists of sixteen, 6 m diameter dish antennas which can be steered to different directions as desired \citep{wu2021tianlai}.

The greatest challenge to all 21 cm intensity mapping experiments is the removal of galactic and extragalactic foreground radiation, which are orders of magnitude brighter than the redshifted 21 cm signal. Fortunately, the foreground radiation has a smooth frequency spectrum, whereas the redshifted 21 cm signal varies stochastically in  frequency. In principle, one can remove the smoothly varying part of the frequency spectrum and recover the 21 cm signal. However, the response functions of radio telescopes are not smooth in either frequency or direction, which will couple with the foreground and generate frequency structure in the data, making the extraction of the 21 cm signal much more difficult. To detect the weak 21 cm signal  requires exquisite knowledge of the instrumental response,  especially the bandpass response and the mutual coupling between different antenna or feed units. Observational results of radio telescopes generally show that there are oscillatory structures in the system bandpass response, and mutual coupling effects between different antennas or feeds. The presence of such structures in radio telescope response has long been known to radio engineers and observers, and it was usually dealt with by a simple procedure of bandpass calibration. However, the precision of such calibration is necessarily limited, and the residues still have a large impact on the high precision observation required by 21 cm cosmology. Therefore, a more quantitative study of the instrument response of radio telescope is needed, and so far there have been only a few such studies (see e.g. \citealt{Kern_2019a,Kern_2019b} for analysis of HERA, and \citealt{li2020tianlai,li2021reflections} for the Tianlai cylinder data).

\begin{figure}[!htbp]
    \centering
	\includegraphics[width=\columnwidth]{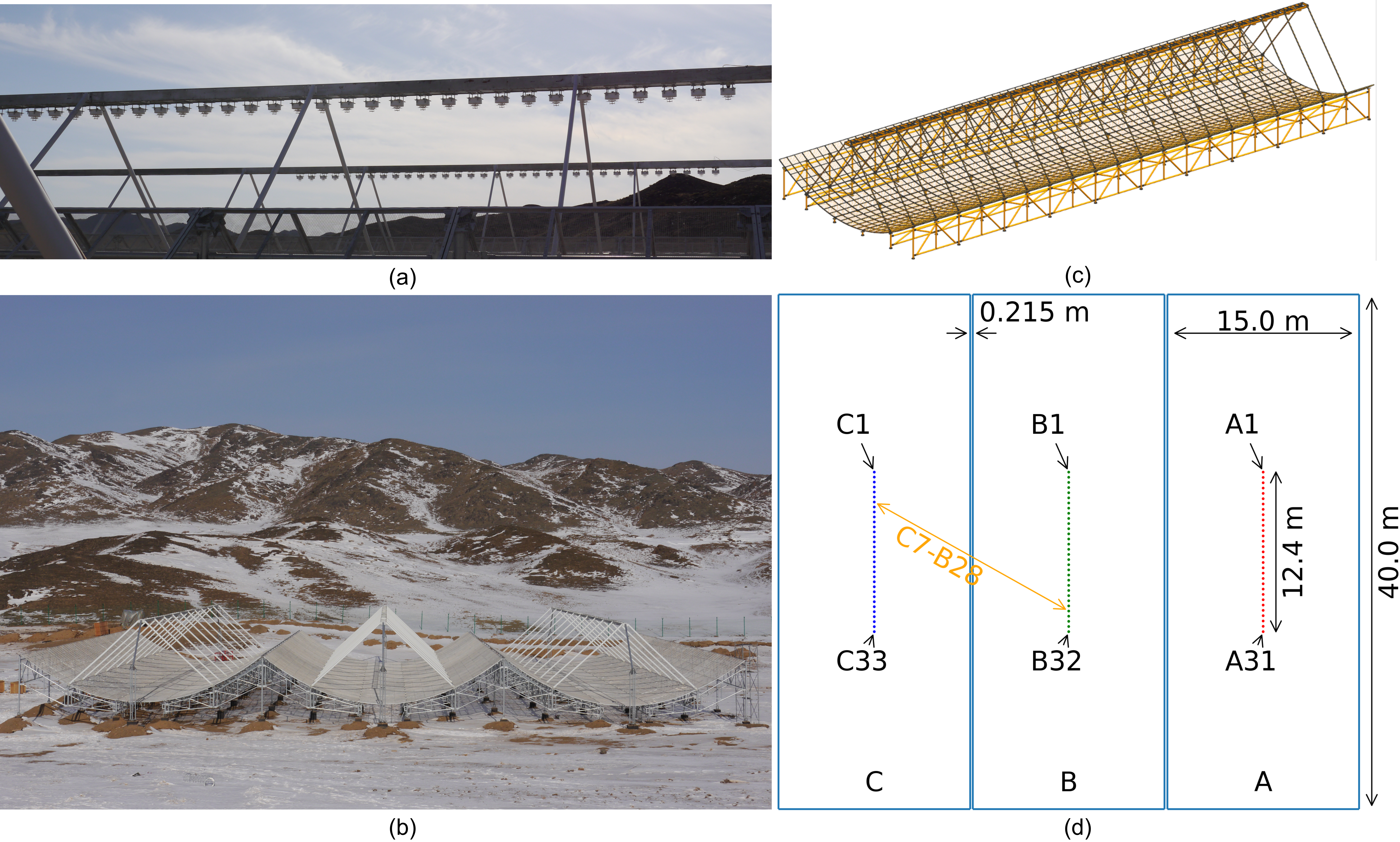}
    \caption{The Tianlai cylinder array configuration. (a): The cylinder array seen from East (feed units visible). (b): The cylinder array seen from South. (c): three-dimensional schematic diagram of one cylinder reflector. (d): feed unit configuration and naming convention.}
    \label{fig:cylinder and feed array configuration}
\end{figure}

In this paper we will make an electromagnetic simulation of the current Tianlai cylinder pathfinder array. At present, such simulations could show some basic  characteristics of the system, which will be helpful for us to interpret the data. Understandably, the simulation results are not completely accurate, due to the many simplifications made in the simulation. We shall compare our simulation results  with experimental data where applicable.

In Fig.~\ref{fig:cylinder and feed array configuration} we show the Tianlai cylinders array configuration, where subfigures (a),(b) show photographs of the telescope from two sides, (c) show a three-dimensional scheme for the structure of one of the cylinder, and (d) show the arrangement of the feed antenna units. Each of the three Tianlai cylindrical reflectors is 40 meters long and 15 meters wide, oriented along the north-south direction and next to each other, with a metal mesh reflecting surface of parabolic cross section, and a focal length of 480 cm \citep{li2020tianlai}. The mesh surface of the cylindrical reflector has been measured at 500 points using a surveyor's theodolite total station; the average RMS (Root Mean Square) deviation is 7.13 mm. A metal, ladder-like ``rung frame'' runs along the focal line, which provide anchors for the receiver feed antennas. As shown in Fig.~\ref{fig:cylinder and feed array configuration}(d), the cylinders are denoted as cylinder A, B, and C from east to west, and there are 31, 32, and 33 feeds installed on the A, B, C cylinders respectively. Unequal numbers of feed units are installed on the three cylinders to break the degeneracy which generates grating lobes in the array beam response. On each cylinder these feeds are evenly distributed with slightly different spacings (41.33 cm, 40.00 cm and 38.75 cm from center to center respectively), so that on all three cylinders the same total length is obtained. In the present work, we shall only consider the A cylinder reflector with 31 feed units; the B and C cylinders may have slightly different results, though qualitatively they would be very similar. We use the same naming convention as described in \cite{li2020tianlai}, with the East-West direction denoted as the Y polarization and the North-South direction as X polarization. The data are labeled in reflector-feed number-polarization format. For example, A1X-A1X denotes the autocorrelation of feed No.1 (the feed at the north end) on reflector A, and A1X-B32X denotes the cross correlation of the X polarization of feed No.1 on reflector A with the X polarization of feed No.32 on reflector B. The instantaneous field of view is a narrow strip along the north-south direction passing through the zenith, limited primarily by the beam of the feed. The designed frequency range of the telescope is 600 MHz - 1420 MHz, corresponding to the redshifts of $0<z<1.4$, with an instantaneous bandwidth of 100 MHz.

The rest of the paper is organized as follows: In Section~\ref{sec:method} we briefly discuss the electromagnetic simulation method for electrically large antennas, then present the simulation and measurement results of the feed for the Tianlai cylinder array in Section~\ref{sec:single feed}. In Section~\ref{sec:feedarray} and Section~\ref{sec:feedarray with reflector} we present simulation results of the feed array and reflector with the feed array respectively in the frequency band of 0.6 GHz - 1.5 GHz, then discuss the performance of the cylinder array, e.g. radiation pattern, peak gain, beam width. In Section~\ref{sec: observation results} we compare the simulation results and observation results in the frequency band of 0.7 GHz - 0.8 GHz, focusing on the bandpass response and mutual coupling effects. Finally, we summarize our results and discuss our conclusions in Section~\ref{sec: conclusions}.

\section{Simulation Method}
\label{sec:method}

We use the electromagnetic (EM) simulation software Ansys HFSS \citep{HFSS} for our simulation.  The vast scales involved in the problem make the simulation 
quite a challenging task. In order to overcome this problem, we use several different kinds of numerical algorithms to make the simulation, each with its own pro and cons. These methods are listed below. 
\begin{itemize}
\item The \textbf{Finite Element Method (FEM)} solves for the electromagnetic fields in the solution region, by setting up meshes over the entire solution volume and solves for the electric field throughout that volume. It provides high accuracy through use of higher order basis functions \citep{jin2015finite}. 

\item The \textbf{Integral Equation (IE)} or \textbf{Method of Moments (MoM)} solver use an integral equation and solves for the currents or equivalent currents on the surface of conducting and dielectric objects. This is often more efficient for problems with large open space, since it does not require establishing grids and solving the field equations in the large surrounding volume \citep{silvestro2010hybrid}. 

\item The \textbf{Finite Element-Boundary Integral (FE-BI)} hybrid method leverages the advantages of the FEM and the IE to achieve the most accurate and robust solution for radiating and scattering problems. In this method, an infinite domain is partitioned into two non-overlapping domains: one bounded FEM domain and one unbounded homogenous exterior region. The coupling of these two domains is taken into account through an appropriate boundary condition at the interface \citep{edgar2011hfss}.

\item The \textbf{Physical Optics (PO)} solver works as part of the IE solver. In PO, a radiation source is used to illuminate the geometry, thus inducing PO currents which then re-radiate. This asymptotic method is extremely useful when solving very large electromagnetic radiation and scattering problems\citep{zhao2018overview}.

\end{itemize}

\begin{figure}[!htbp]
    \centering
	\includegraphics[width=0.8\columnwidth]{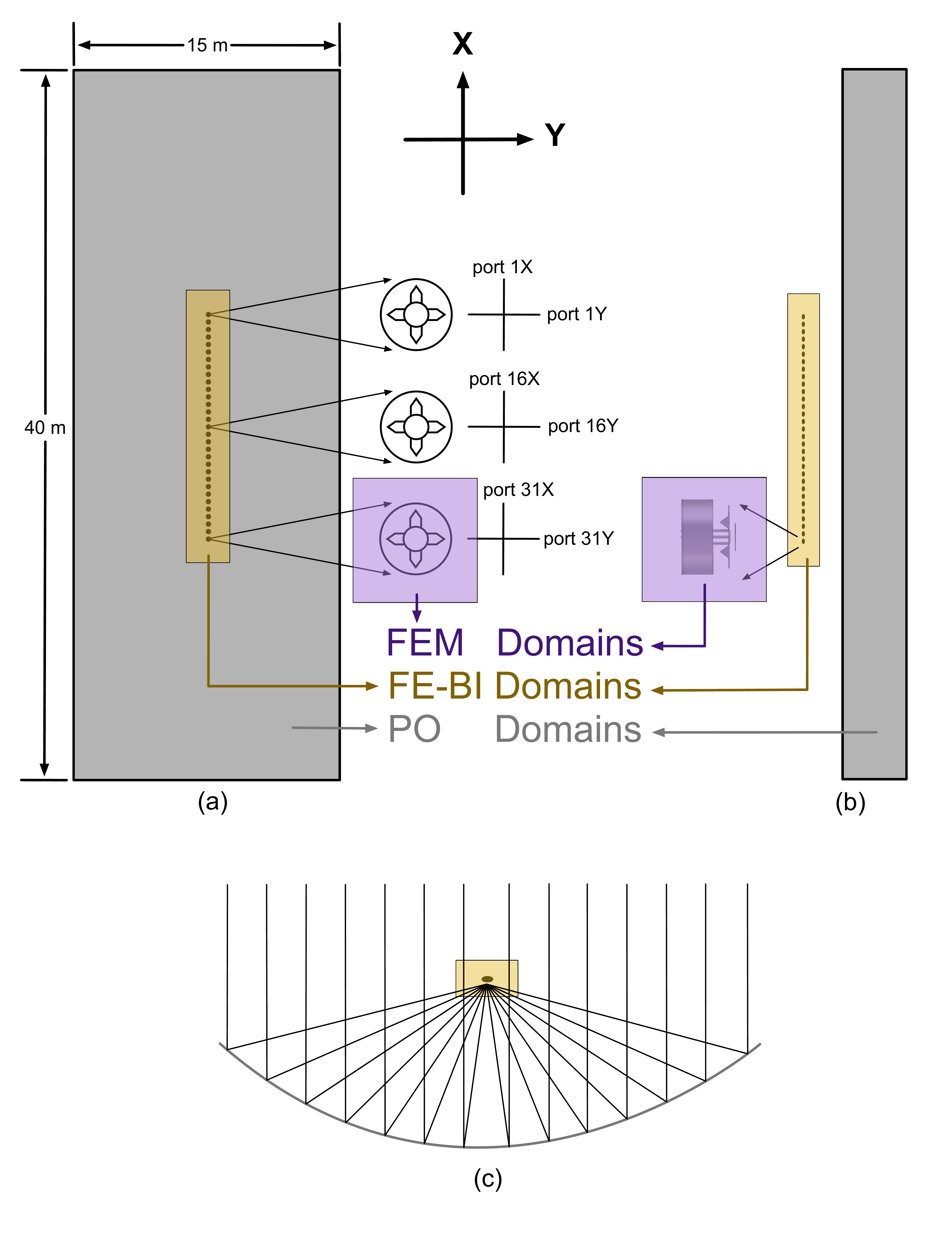}
    \caption{The Tianlai cylindrical reflector with feed array simulation models. (a): top view. (b): side view. (c): front view (with light rays). The X direction is along the cylinder, Y direction is along the paraboloid. The FEM domains (in purple color) for the single feed model, the FE-BI domain (in yellow color) for the feed array model, and the PO (in silver color) for the cylindrical reflector model are shown in the figure. Three feeds in the top view are enlarged to show the port definitions.}
    \label{fig:simulation model}
\end{figure}

To simulate an antenna of large size and complicated geometry such as the Tianlai cylinder array, it is necessary to use a combination of methods to achieve the necessary precision with practically available computing resources, such as the amount of computation and the size of memory and data storage. 
We make the simulation in three steps: (1) simulation of a single feed, (2) simulation of
an array of feed units, and (3) simulation of the feed array and the cylindrical reflector, as depicted in  Fig.~\ref{fig:simulation model}. 
We choose the simulation method of each step according to the size of the problem. For the single feed model (highlighted with purple color in Fig.~\ref{fig:simulation model}), which has complicated structure within the scale of a wavelength, the FEM solver is used. For the feed array model (highlighted with yellow color), which includes 31 feeds, solution with the FEM solver becomes impractical. Instead, the FE-BI solver is used. In this set-up, a boundary is assigned on the box enclosing the feed array, with the radiation at the boundary given by the integral equations. For the cylindrical reflector (highlighted in silver color), as the spatial structure is larger than $\sim$200 wavelengths in dimension,  the PO solver is used, which can provide an accurate approximation with  significantly reduced computational resource requirements. The feed array is used as a radiation source to illuminate the reflector, inducing PO currents that then re-radiate. For a server equipped with 2.2 GHz, 20 cores CPU and 128 GB RAM, about 60 hours are needed to ``mesh'' the model, and after that 2 hours to solve for each frequency point. In order to obtain the bandpass curve, we computed about 300 frequency points; the total computing time is about 30 days.

In our simulation, we shall investigate the beam pattern from a feed, as well as the coupling effects between different feeds. For the latter, 
the linear response of an electronic system can be described by the scattering matrix or S-parameters, 

\begin{equation}
    \rm S_{\rm ij}=\frac{U_{\rm i}^{rec}}{U_{\rm j}^{in}}
    \label{eq:S-parameters define}
\end{equation}

\noindent where $\rm U_{\rm i}^{rec}$ represents the voltage received from port $i$, and $\rm U_{\rm j}^{in}$ represents the voltage incident on port $j$. 
The $\rm S_{ii}$ are the reflection coefficients, and $\rm S_{ij} (i\neq j)$ the transmission coefficients. For an antenna or a feed, the reflection coefficients indicate impedance match characteristics, and the transmission coefficients indicate mutual coupling characteristics of the given two ports.

\section{Single Feed}
\label{sec:single feed}

\begin{figure}[!htbp]
    \centering
	\includegraphics[width=0.8\columnwidth]{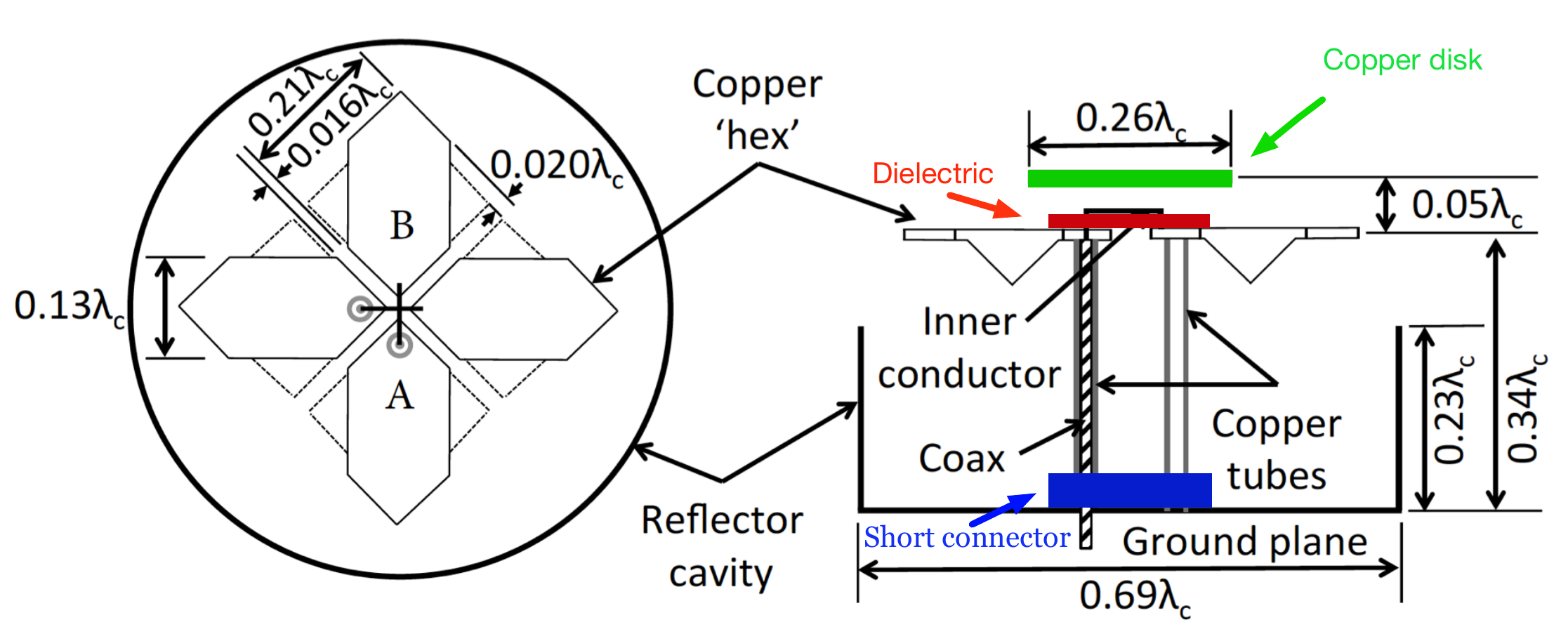}
    \caption{The structure of a feed used in the Tianlai cylinder array, as shown by the top view (left) and side view (right). 
    }
    \label{fig:feed schematic}
\end{figure}

The feed antennas for the Tianlai cylindrical telescope are designed and optimized to achieve good performance over the frequency range of 600 MHz - 1420 MHz, corresponding to the redshifts of $0<z<1.4$. The feed structure is shown in Fig.~\ref{fig:feed schematic}. The design has evolved from a planar `four-square' dipole into a `four-hex' shape dipole by folding the corners of each of the squares toward the ground plane. This folding is found to reduce cross coupling between polarizations, and to improve the impedance match of the feed. A `coffee can' reflector cavity structure is also added to house the feed. The 
design of this feed has been described in detail in \citet{liu2014design} and \citet{cianciara2017simulation}. In the eventual product which has been installed on the array, some small structures are added to improve its performance; these modifications are shown in color in Fig.~\ref{fig:feed schematic}. A short connector (in blue color) is added to improve the impedance match characteristics. A dielectric (in red color) is used for reducing the cross-coupling between the two polarization radiation blades. The height of the copper disk (in green color) is adjusted to create a resonant structure which increases the gain.

\begin{figure}
    \centering
	\includegraphics[width=0.8\columnwidth]{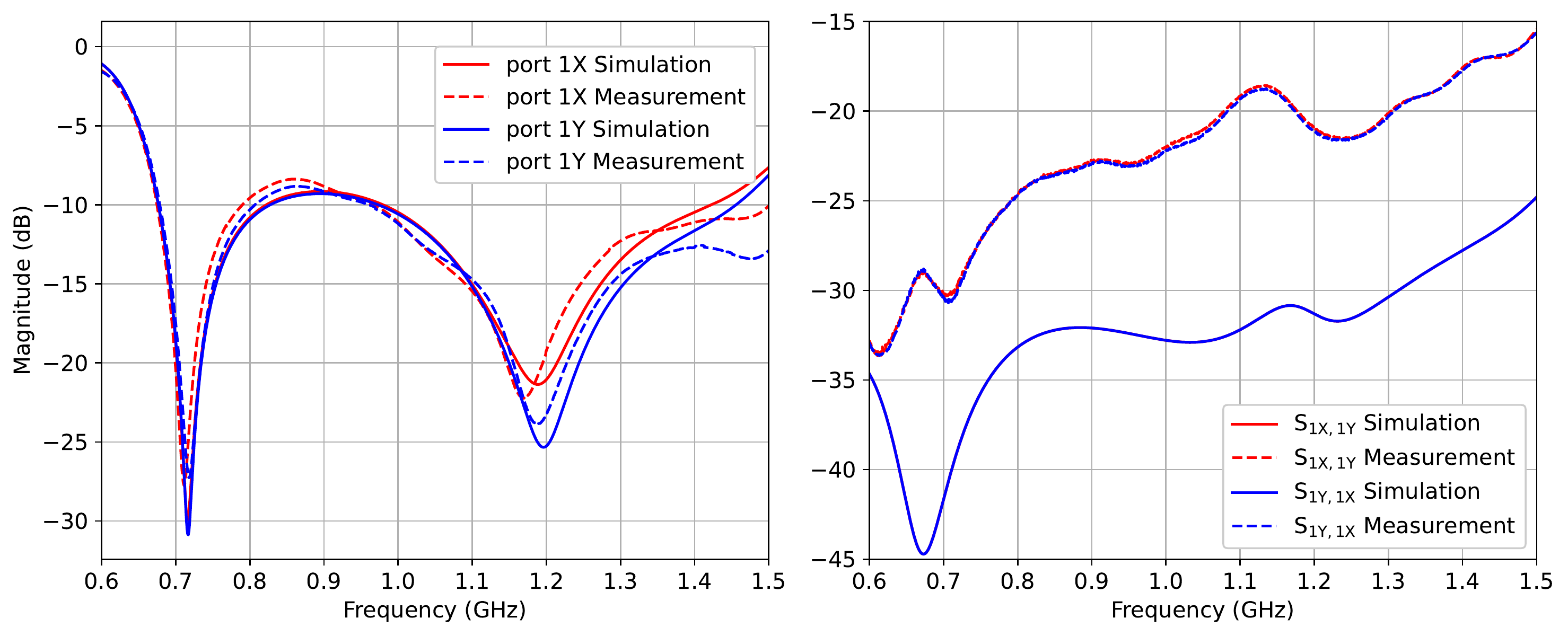}
    \caption{Left: simulated and measured magnitude of the reflection coefficients ($\rm S_{1X,1X}$ and $\rm S_{1Y,1Y}$) of the two polarizations of the feed. Right: simulated and measured magnitude of the transmission coefficients ($\rm S_{1X,1Y}$ and $\rm S_{1Y,1X}$) of the two polarizations of the feed. }
    \label{fig:s11 s21 measurement and simulation}
\end{figure}

The simulated reflection coefficients ($\rm S_{ii}$) and transmission coefficients ($\rm S_{ij}$) are shown in Fig.~\ref{fig:s11 s21 measurement and simulation}, along with actual measurement result obtained in a microwave anechoic chamber with a two-port Vector Network Analyzer (VNA). As can be seen from Fig.~\ref{fig:s11 s21 measurement and simulation},  the magnitude of the reflection coefficients varies with frequency, but it is well below -10 dB in the band of 0.7 GHz - 1.4 GHz, indicating that the two ports are well matched to the standard 50 $\Omega$ input impedance of the receiver, and the difference between simulation and measurement is generally small over the whole band. The reflection coefficients of the two polarizations are slightly different due to the small asymmetry in the structure of the radiation blades, where the inner conductor of 1X is a little taller than that of 1Y. The difference between the simulation and measurement for transmission coefficients is larger, though the general shapes of the curves are similar. The actual mutual coupling between the two polarizations as indicated by the transmission coefficients is larger than the simulation, probably due to small, un-modeled details in the structure and the manufacturing precision. 

\begin{figure}
    \centering
	\includegraphics[width=\columnwidth]{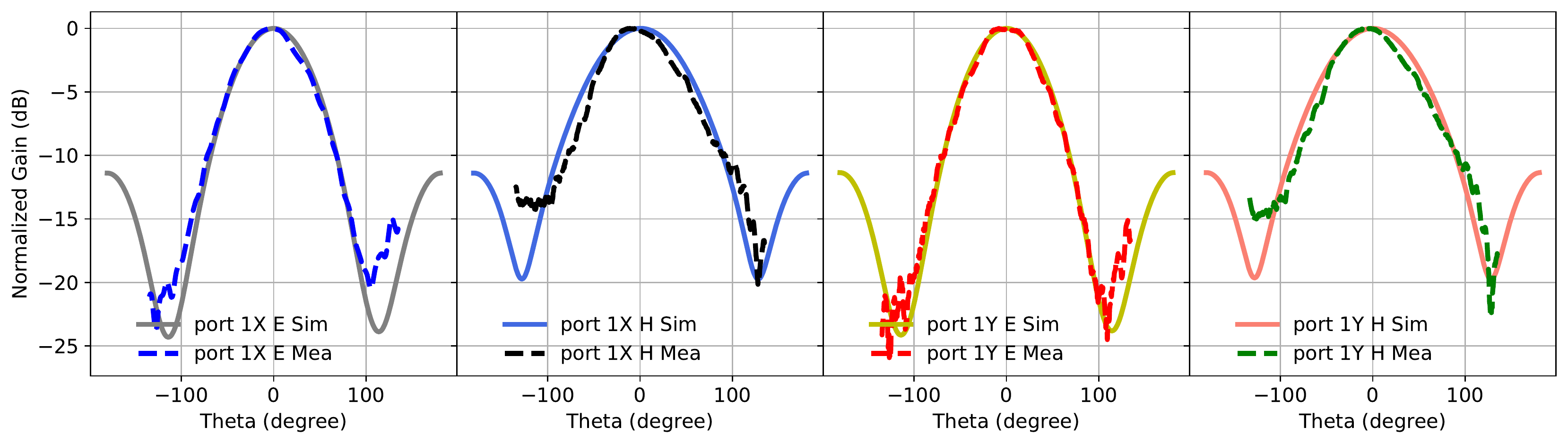}
	\includegraphics[width=\columnwidth]{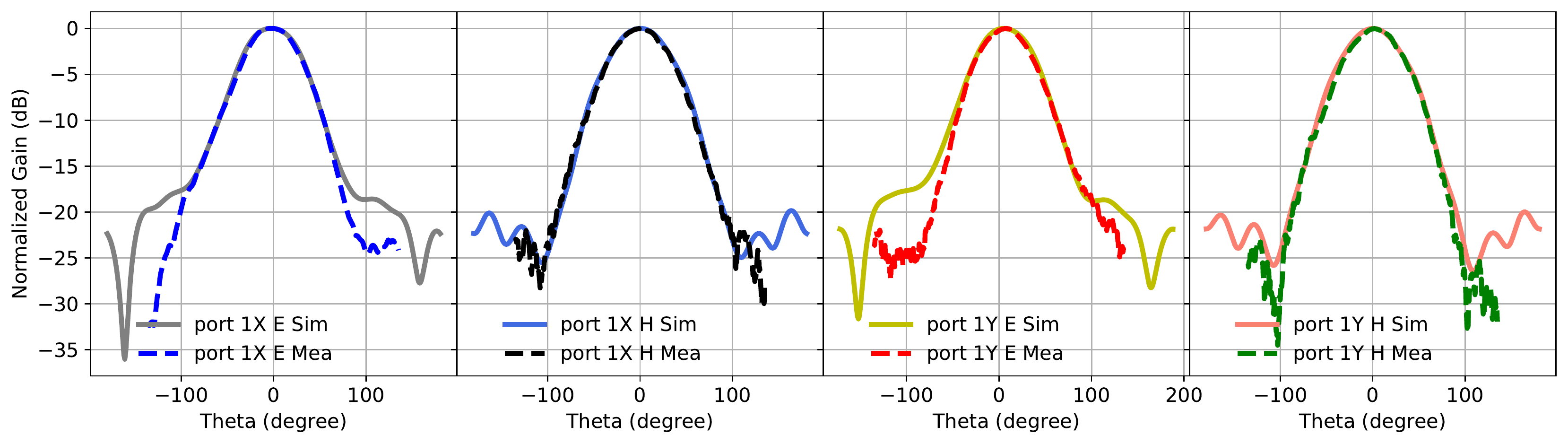}
    \caption{Measured and simulated radiation pattern of a single feed. Top: 0.7 GHz, Bottom: 1.4 GHz. }
    \label{fig:single_feed_pattern}
\end{figure}

The radiation pattern of the feed is measured in the same microwave anechoic chamber. The normalized radiation pattern at 0.7 GHz and 1.4 GHz are shown in Fig.\ref{fig:single_feed_pattern}. For each polarization, the direction of the E field is defined as the E plane, and the H plane is the plane perpendicular to it. 
The feed has good symmetrical patterns in the main beam, and reasonably small sidelobes. There is good agreement between the measured beam and the simulated beam in the range $-135^{\circ} < \theta <135^{\circ}$ where the measurement has been taken, giving confidence to the validity of the simulation. 

\begin{table}[!htbp]
    \caption{Beam width (-10 dB) of Port 1X and Port 1Y}
    \label{tab:feed beam pattern -10dB width}
    \centering
    \footnotesize
    \setlength{\tabcolsep}{4pt}
    \renewcommand{\arraystretch}{1.2}
    \begin{tabular}{lcccccccc}
        \hline
        frequency (GHz) & $0.7$ & $0.8$ & $0.9$ & $1.0$ & $1.1$ & $1.2$ & $1.3$ & $1.4$\\
        \hline
        Port 1X (E plane) & $135.87^\circ$ & $144.25^\circ$ & $148.19^\circ$ & $144.84^\circ$ & $135.82^\circ$ & $128.52^\circ$ & $120.22^\circ$ & $113.17^\circ$\\
        \hline
        Port 1X (H plane) & $171.39^\circ$ & $178.19^\circ$ & $177.71^\circ$ & $175.87^\circ$ & $174.67^\circ$ & $166.31^\circ$ & $151.70^\circ$ & $136.27^\circ$\\
        \hline
        Port 1Y (E plane) & $135.57^\circ$ & $143.74^\circ$ & $147.76^\circ$ & $144.86^\circ$ & $136.42^\circ$ & $128.79^\circ$ & $120.61^\circ$ & $113.01^\circ$\\
        \hline
        Port 1Y (H plane) & $178.66^\circ$ & $178.48^\circ$ & $176.73^\circ$ & $174.43^\circ$ & $165.13^\circ$ & $151.00^\circ$ & $135.47^\circ$ & $120.47^\circ$\\
        \hline
    \end{tabular}
\end{table}

Table~\ref{tab:feed beam pattern -10dB width} shows the simulated E-plane and the H-plane -10 dB beam width for the two polarizations of the feed. The 
H-plane beam width is wider than the E-plane one. The beam width becomes narrower at higher frequency. Simulation and measurement results also show that the gain and Half Power Beam Width (HPBW) of the feed are weakly frequency-dependent.

\section{Feed Array}
\label{sec:feedarray}

In the above we simulated an individual feed, but in each of the Tianlai cylinders, an array of feed units are installed along the focus line. These feeds are densely distributed within the near field of each other, thus the performance of each feed is not completely independent, and mutual coupling effects could be important. In this section, we first construct a feed array simulation model based on the single feed model. Here we focus on the variation of performance for one feed in different positions within the feed array, comparing with the simulation results for the single feed in isolation. In the next section we will construct a model including additionally the cylindrical reflector.

\begin{figure}[!htbp]
    \centering
	\includegraphics[width=\columnwidth]{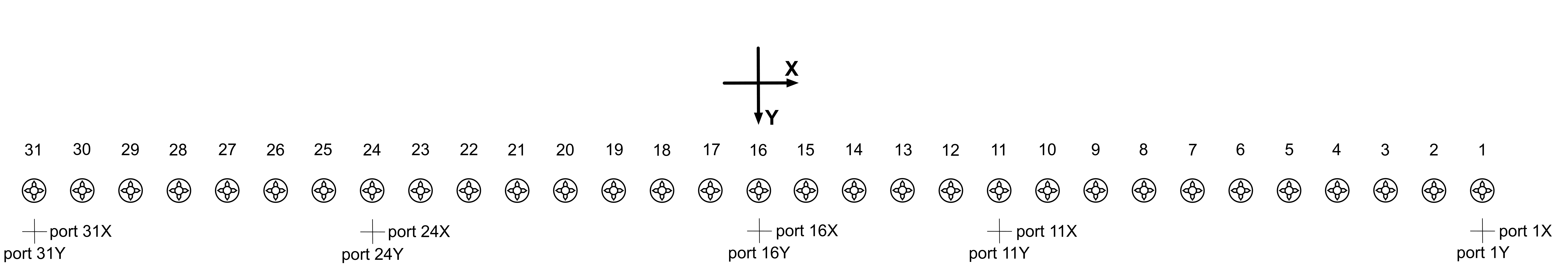}
    \caption{Feed array simulation model. }
    \label{fig:31 feeds model}
\end{figure}

The feed array model is shown in Fig.~\ref{fig:31 feeds model}. The model includes 31 evenly distributed feeds, with a center-to-center spacing of 38.75 cm. A simplification here is we neglected the rung frame used to hang the feeds, and the fiberglass struts that support the rung frame from the cylinder itself.

\begin{figure}
    \centering
	\includegraphics[width=0.8\columnwidth]{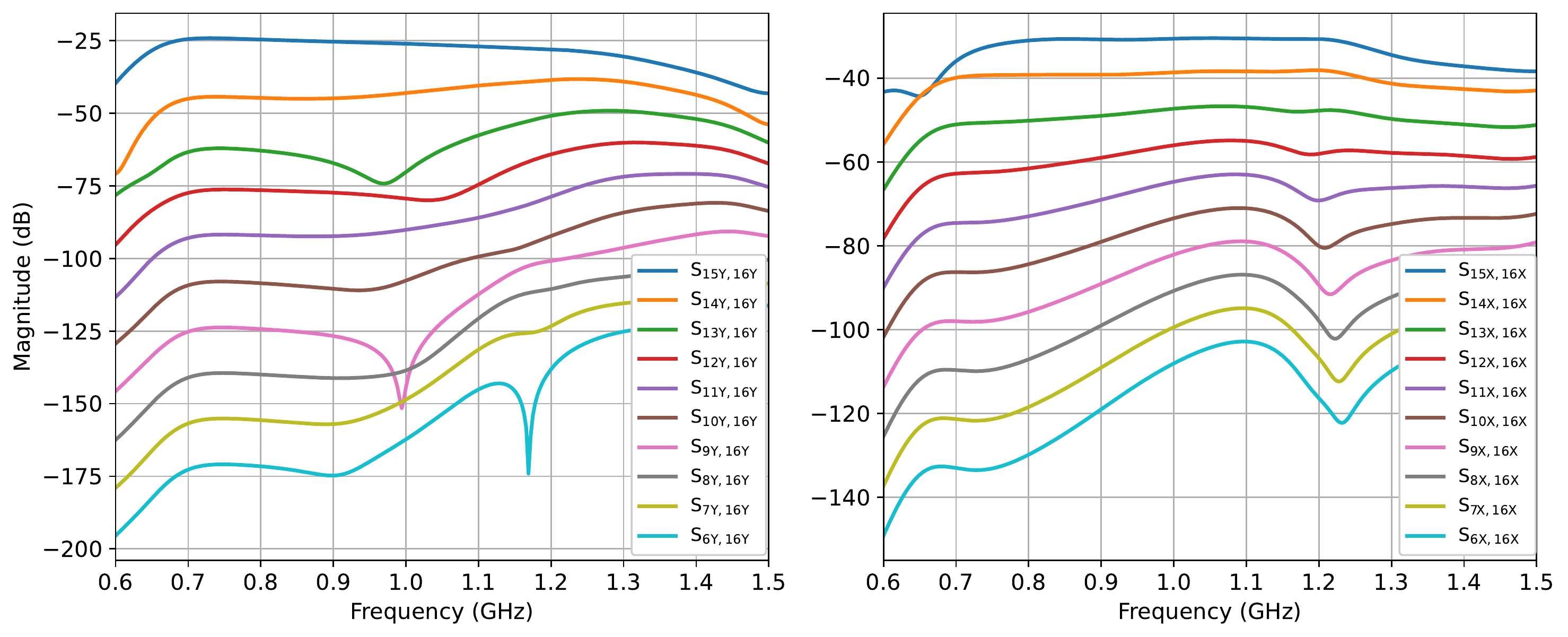}
    \caption{Simulated transmission coefficients of the feed array. Left: Y polarization ports. Right: X polarization ports. }
    \label{fig:feed array S parameters mutual coupling}
\end{figure}

\begin{figure}[!htbp]
    \centering
	\includegraphics[width=0.6\columnwidth]{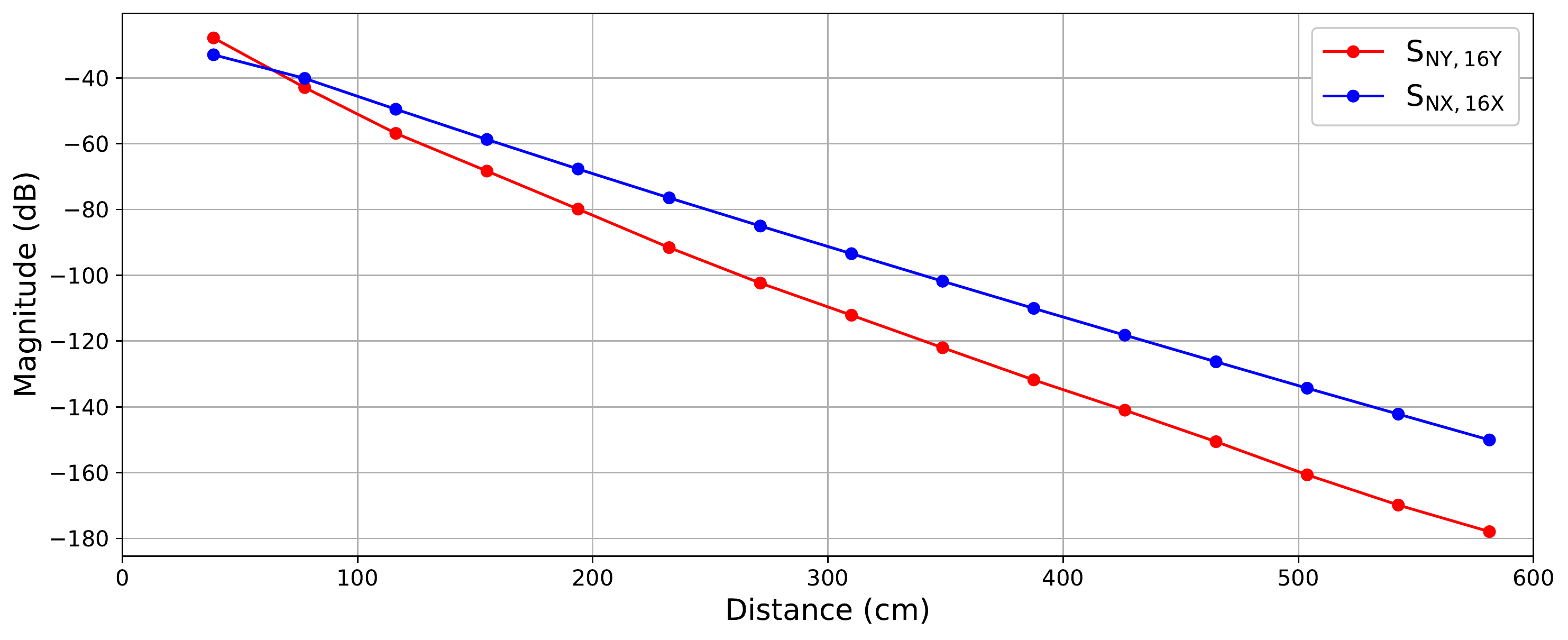}
    \caption{Mean magnitude and of the transmission coefficients averaged over the frequency  range of 0.6 GHz - 1.5 GHz as a function of the distance between feeds. }
    \label{fig:31 feeds mutual mean}
\end{figure}

As previously described, the feed array simulation model is enclosed by an FE-BI boundary, and an integral equation solver is used at the radiation boundary. We compared the reflection coefficients of each feed within the array to the single feed model, and there is only a slight difference. This result confirms that the feed array structure has no impact on the reflection characteristics of the feed.

Fig.~\ref{fig:feed array S parameters mutual coupling} shows the simulated S-parameter of the feed array, and Fig.~\ref{fig:31 feeds mutual mean} shows the mean magnitude of the S-parameters with respect to frequency over 0.6 GHz - 1.5 GHz. For both X polarization ports and Y polarization ports the transmission coefficients decrease linearly with the distance between ports. Results show that for the Y polarization, the S-parameter,  i.e. the crosstalk between the adjacent two Y polarized ports (e.g. $\rm S_{2Y,1Y}$), is around -25 dB. The feeds further away have smaller coupling; for the feed which is 10 spacing units away, it drops to  -170 dB.  Similar results are found for the X polarization ports.  The S-parameter between the adjacent two feed units is also around -25 dB, but decreases as the distance increases, though the drop is not as steep as the Y-polarization. At 10 spacing units, it drops to around -120 dB. These couplings are quite small, but we shall see below that the couplings are significantly larger with the reflector. 

\begin{figure}
    \centering
	\includegraphics[width=0.8\columnwidth]{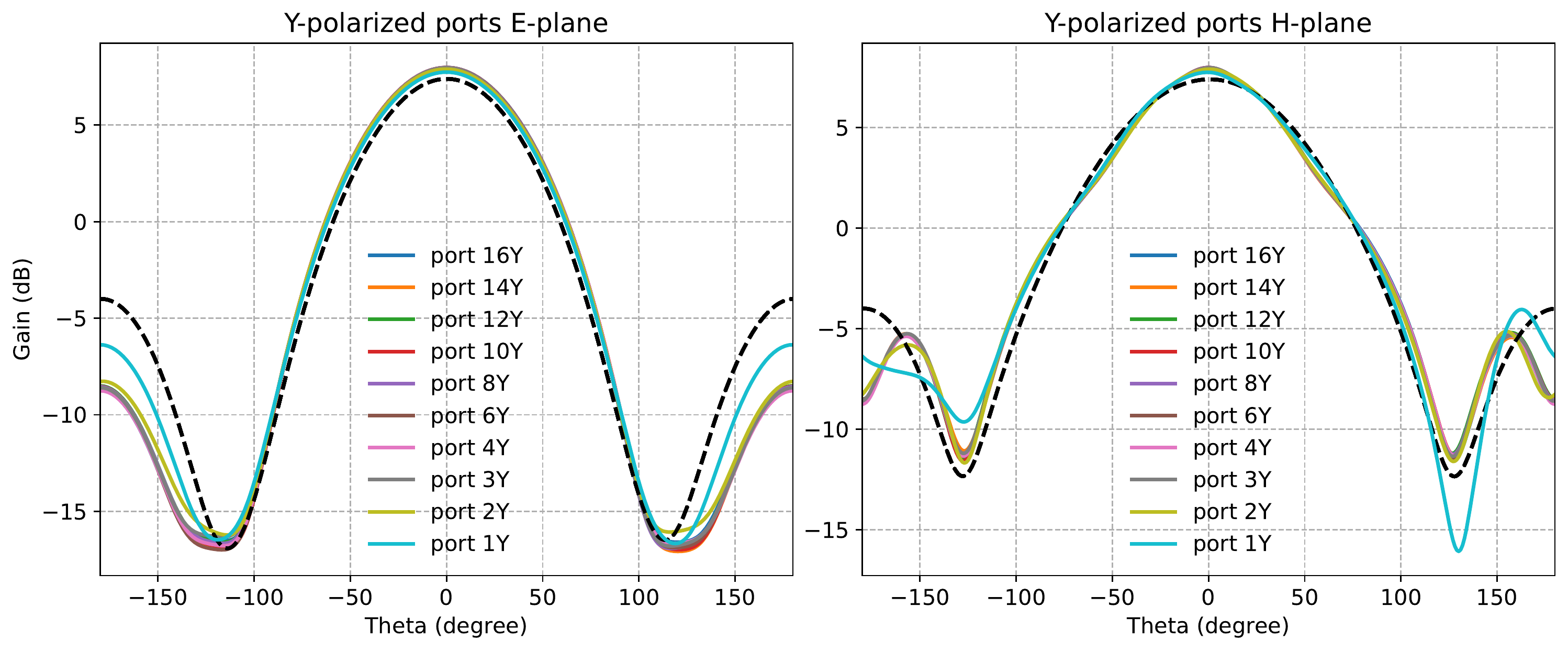}
	\includegraphics[width=0.8\columnwidth]{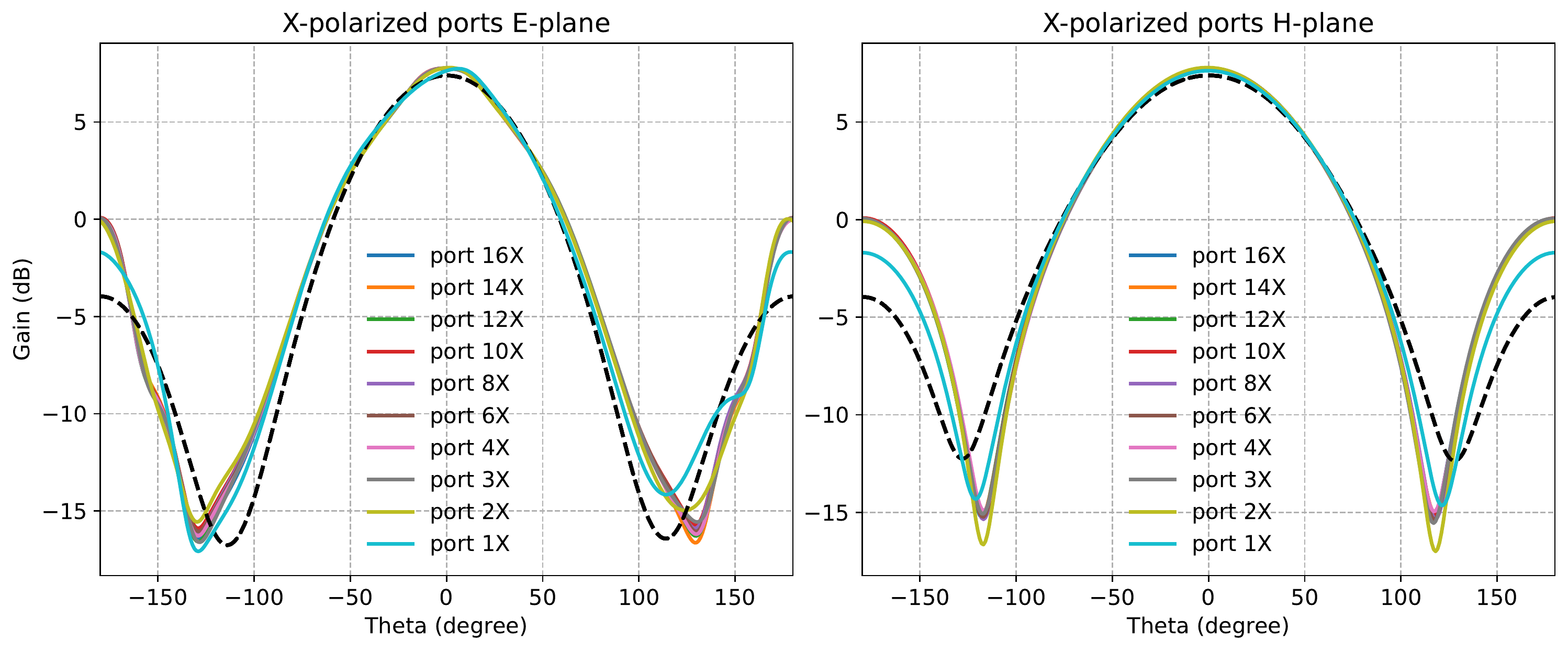}
    \caption{Far-field radiation pattern of the feeds in the array at 0.7 GHz, the radiation pattern of a single feed is shown as black dashes. Top panels: Y-polarization feeds, Bottom panels: X-polarization feeds. Left: E plane. Right: H plane.}
    \label{fig:31feeds radiation pattern ports E and H plane 0.7G}
\end{figure}

Fig.~\ref{fig:31feeds radiation pattern ports E and H plane 0.7G} shows the simulated far field radiation pattern of several ports in the feed array at 0.7 GHz. As a comparison, the radiation pattern of the single feed model is also plotted with bold black dashes. For the Y-polarization (top panels), The left (right) panel shows the beams on the E (H) plane. 
For the Y polarization ports, the radiation patterns are all symmetric w.r.t. $\theta=0^\circ$. The main lobes in the feed array are basically the same as that of the single feed model, but the differences in the side lobes are apparent. The beam pattern is not symmetrical w.r.t. $\theta=0^\circ$ in the H plane, especially for the feed in the edge position, though the main lobe is still basically the same as that of the single feed model. For the X-polarization (bottom panels), the H plane beams are symmetric w.r.t. $\theta=0^\circ$ but the E plane beams not. Again, the main lobes in the feed array are basically the same as that of the single feed model, but there are differences in the side lobes. These simulation results show that for the feed elements in feed array, their radiation characteristics are to a large extent consistent with the single feed model, except for those feeds near the edge, where the angular asymmetry is more apparent.

\section{Feed Array with Reflector}
\label{sec:feedarray with reflector}

Finally, we perform electromagnetic co-simulations of the feed array at the primary focus of the cylindrical reflector, as depicted in Fig.~\ref{fig:simulation model}. A simplified model for the reflector is used: we assumed an ideal parabolic cylinder without supporting frame structure, and the reflecting surface is assumed to be a perfect electrical conductor with no thickness. For reflector antennas the recommended ratio of power at the reflector edge to the centre is -10 dB. In our case, the angle spanned by the cylinder and center is around $152^\circ$, which is comparable with the -10 dB beam width of the feed given in Table.~\ref{tab:feed beam pattern -10dB width}.

\subsection{Reflection and transmission characteristics}
\label{subsubsection: Reflection and transmission characteristics}

\begin{figure}[htbp]
    \centering
	\includegraphics[width=0.8\columnwidth]{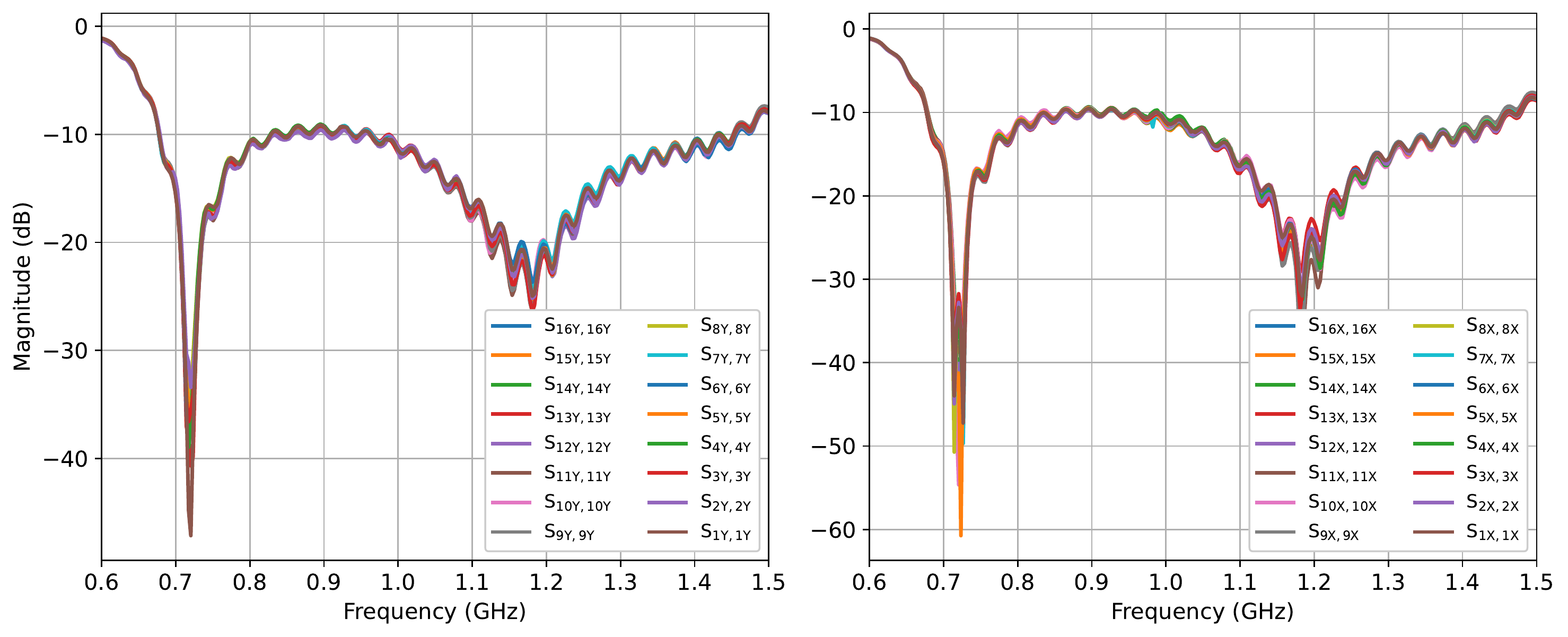}
    \caption{Simulated reflection coefficients of the feed array with the cylinder reflector model. Left: Y-polarized ports. Right: X-polarized ports.}
    \label{fig:reflector with feed array S parameters self coupling}
\end{figure}

Fig.~\ref{fig:reflector with feed array S parameters self coupling} shows the reflection coefficients for all X-polarized ports and Y-polarized ports, to be compared with the single feed model in Fig.~\ref{fig:s11 s21 measurement and simulation}. An oscillatory feature appears in the frequency response, which we interpret as arising from the standing wave resonance between the feeds and the reflector. The designed focal length of the reflector is 4.8 meter. An RF signal radiated from the feed into free space, then reflected by the cylindrical reflector surface and re-coupled to the feed travels a distance of 9.6 m. The cycle period in the frequency response is around 31.1 MHz, corresponding to a wavelength of $\sim 9.646$ m.

\begin{figure}[!htbp]
    \centering
	\includegraphics[width=0.9\columnwidth]{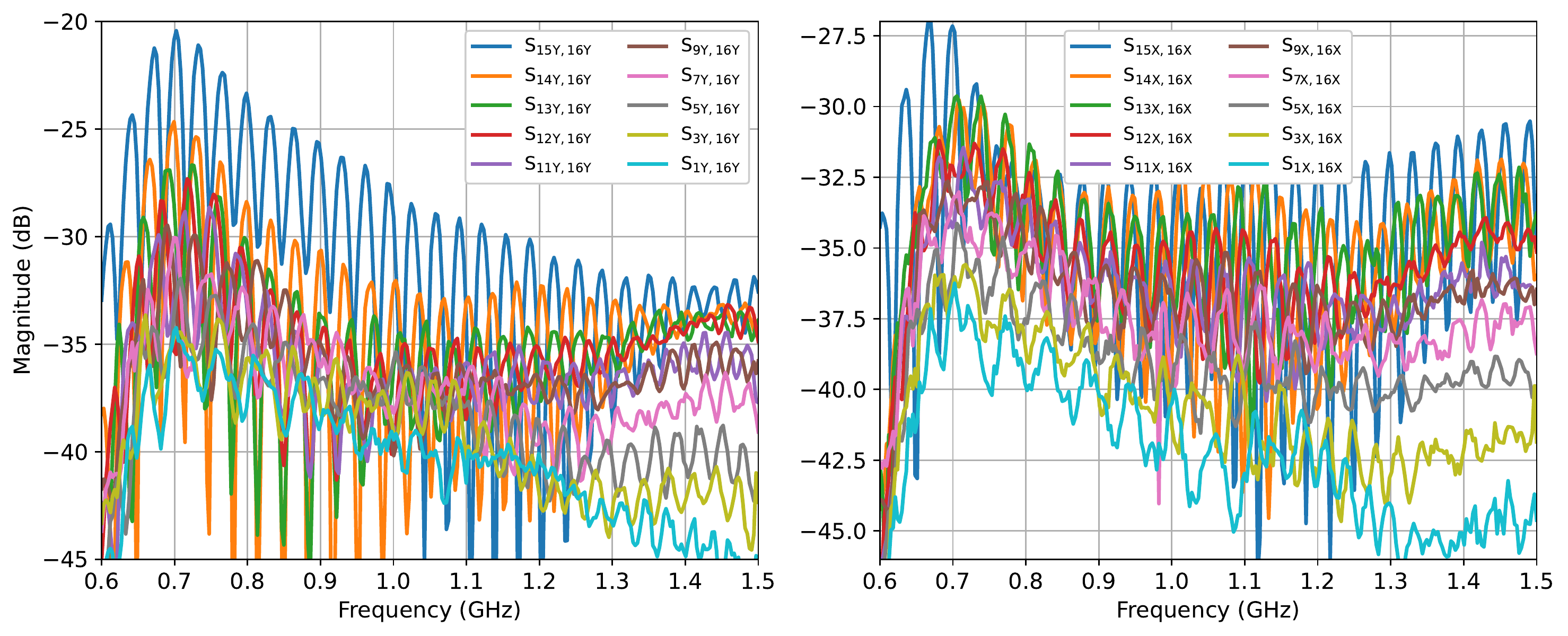}\\
	\includegraphics[width=0.9\columnwidth]{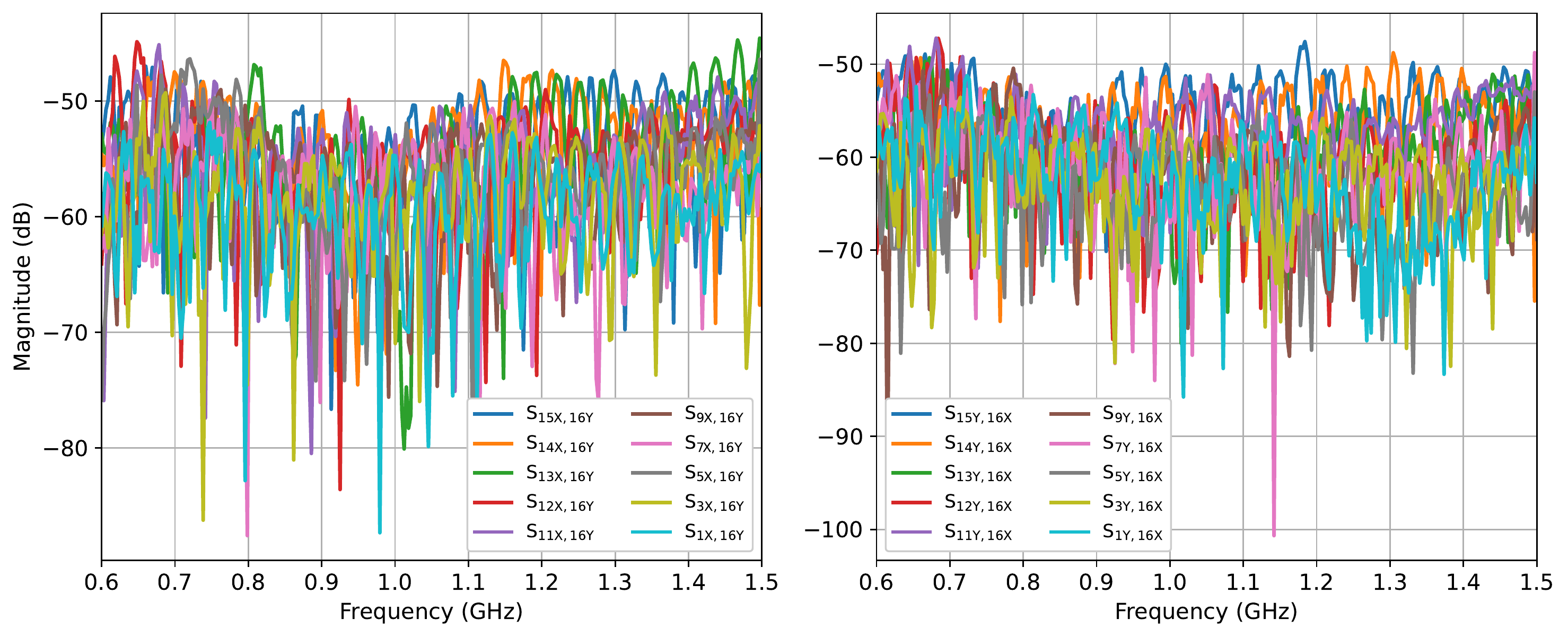}
    \caption{Simulated transmission coefficients of the feed array with the cylinder reflector model. Top Left: YY, Top Right: XX, Bottom Left: XY, Bottom Right: YX.}
    \label{fig:reflector with feed array S parameters mutual coupling}
\end{figure}

\begin{figure}[htbp]
    \centering
	\includegraphics[width=0.7\columnwidth]{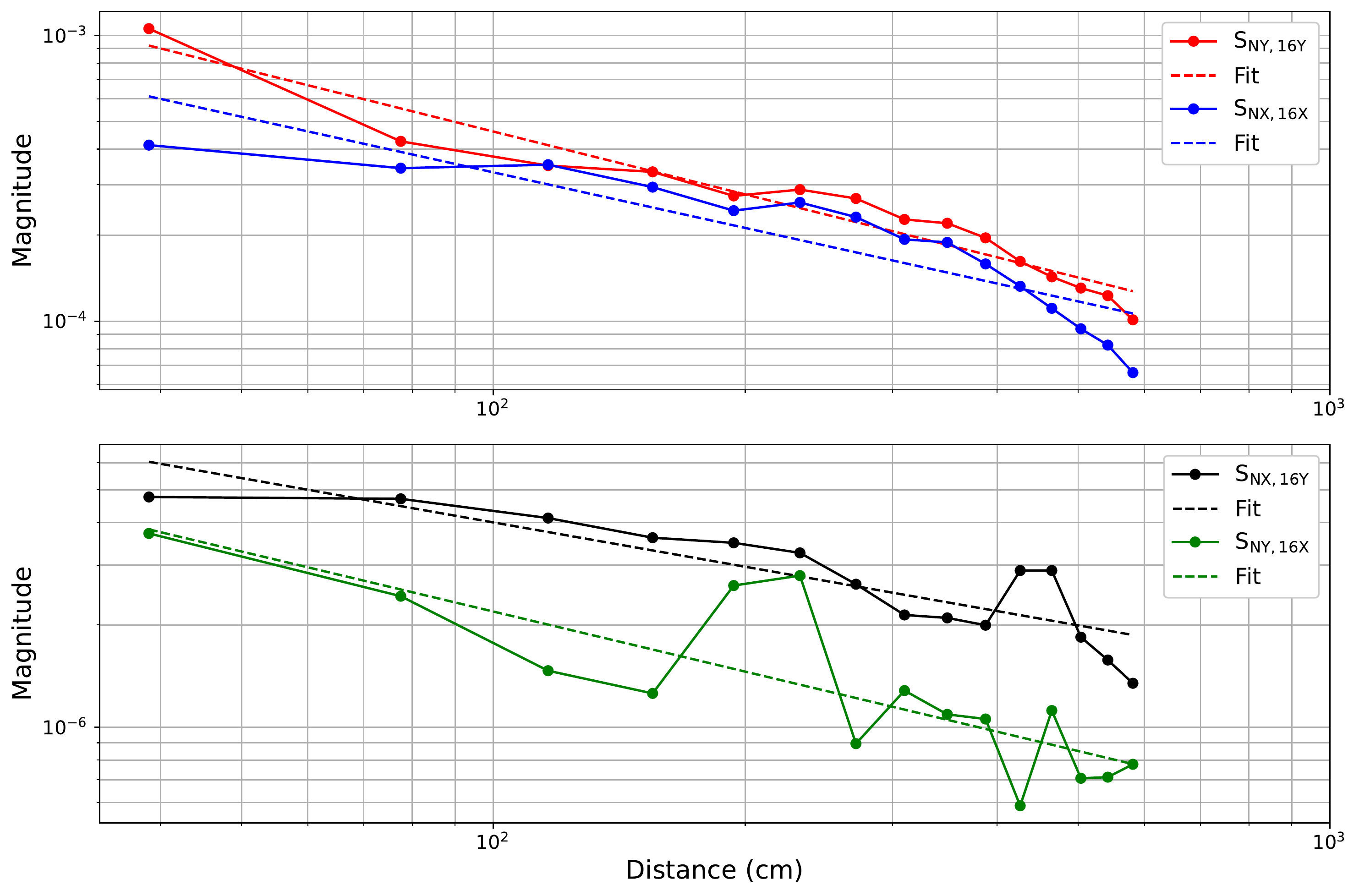}
    \caption{Mean magnitude of the transmission coefficients averaged over 0.6 GHz - 1.5 GHz as a function of the distance between the feed. We also plot a fitting line for the S-parameters vs distance, the best fits are: $S_{\rm NY,16Y}\sim D^{-0.73}$, $S_{\rm NX,16X}\sim D^{-0.65}$, $S_{\rm NX,16Y}\sim D^{-0.43}$ and $S_{\rm NY,16X}\sim D^{-0.59}$.}
    \label{fig:cylinder with feedarray S parameters mutual coupling mean}
\end{figure}

Fig.~\ref{fig:reflector with feed array S parameters mutual coupling} shows the simulated transmission coefficients of the feed array with the reflector. 
The oscillatory features appear in the transmission coefficients with the same interval as in the reflection coefficients. Comparing with the mutual coupling results for the feed-array-only model, the coefficients are larger for the model with the reflector, especially for the port pairs with large separations. This is expected, as the electromagnetic waves can be coupled from one feed to another by reflections from the cylinder surface.

The top panels of Fig.~\ref{fig:reflector with feed array S parameters mutual coupling} show the coupling for pairs of the same polarization. For the center feed and its adjacent one, the S-parameter for the Y-polarization could reach almost -20 dB, and the average over 0.6 GHz - 1.5 GHz is around -30 dB. The coupling to the next one is about 5 dB smaller, and between the center feed and the edge feed the peak/average drops to around -35 dB /-40 dB. The cross-couplings between the X-polarizations are smaller, which is expected as the feeds are aligned in the H-plane, with a peak of about -27 dB for the adjacent ones, and decreases with increasing distance. 

The bottom panels of Fig.~\ref{fig:reflector with feed array S parameters mutual coupling} show the simulated transmission coefficients for a pair of  orthogonal polarizations. Results show that these transmission coefficients are below -50 dB and have similar magnitudes. The more distant pairs are not much smaller in magnitude, perhaps because the direct coupling is already very small, and the indirect couplings are more complicated and do not necessarily decrease much over this distance. 

These overall trends of coupling dependence on feed pair distance is also shown in Fig.~\ref{fig:cylinder with feedarray S parameters mutual coupling mean}, which shows the mean magnitude of the S-parameters averaged over the whole frequency range of 0.6 GHz - 1.5 GHz. Fitting the relation with power law relation, we find the best fit parameters are $S_{\rm NY,16Y}\sim D^{-0.73}$, $S_{\rm NX,16X}\sim D^{-0.65}$, $S_{\rm NX,16Y}\sim D^{-0.43}$ and $S_{\rm NY,16X}\sim D^{-0.59}$. The scatter in the relation is relatively large, so we could not draw very strong conclusions on these relations, but the general trend is that the cross-couplings decrease with the distance between the feeds, and the same-polarization cross-couplings decrease  more rapidly than the orthogonal cross-couplings.

\subsection{Radiation pattern and Bandpass}

\begin{figure}[!htbp]
    \centering
	\includegraphics[width=0.85\columnwidth]{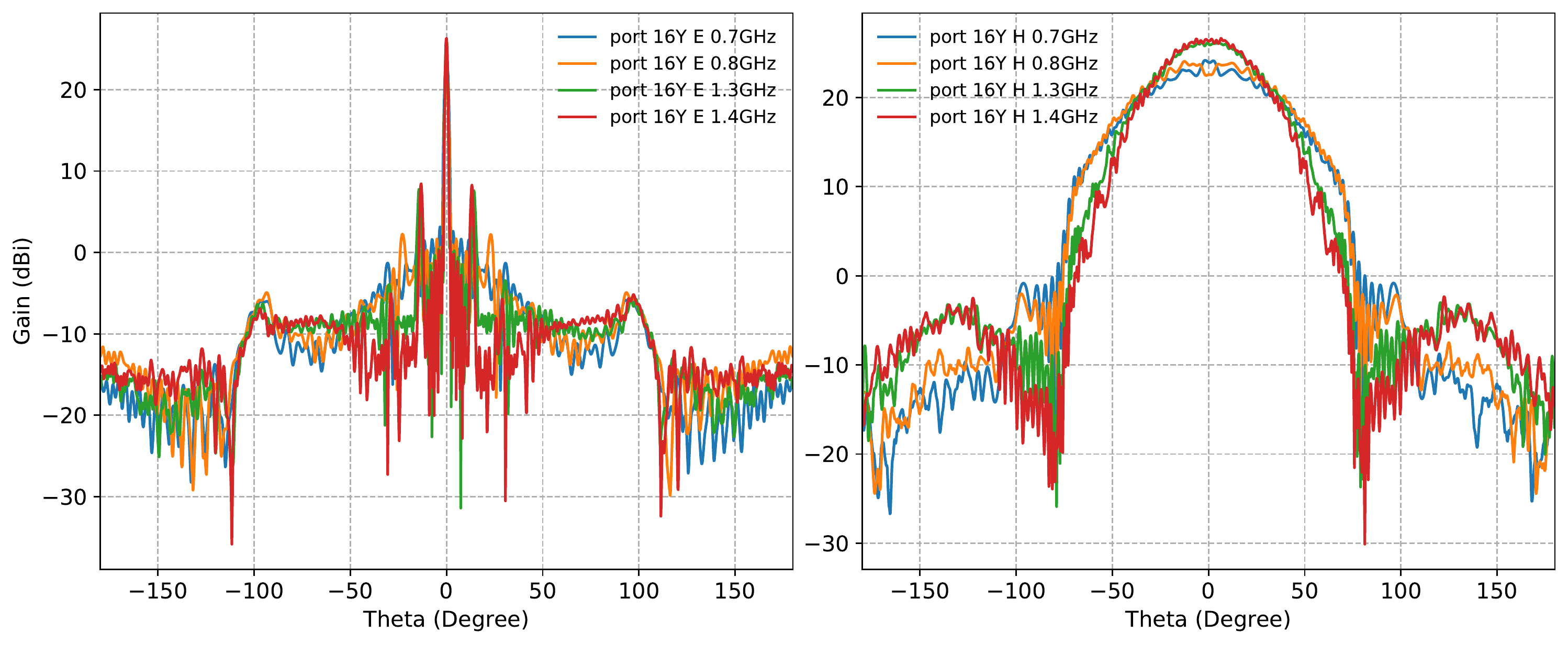}
	\includegraphics[width=0.85\columnwidth]{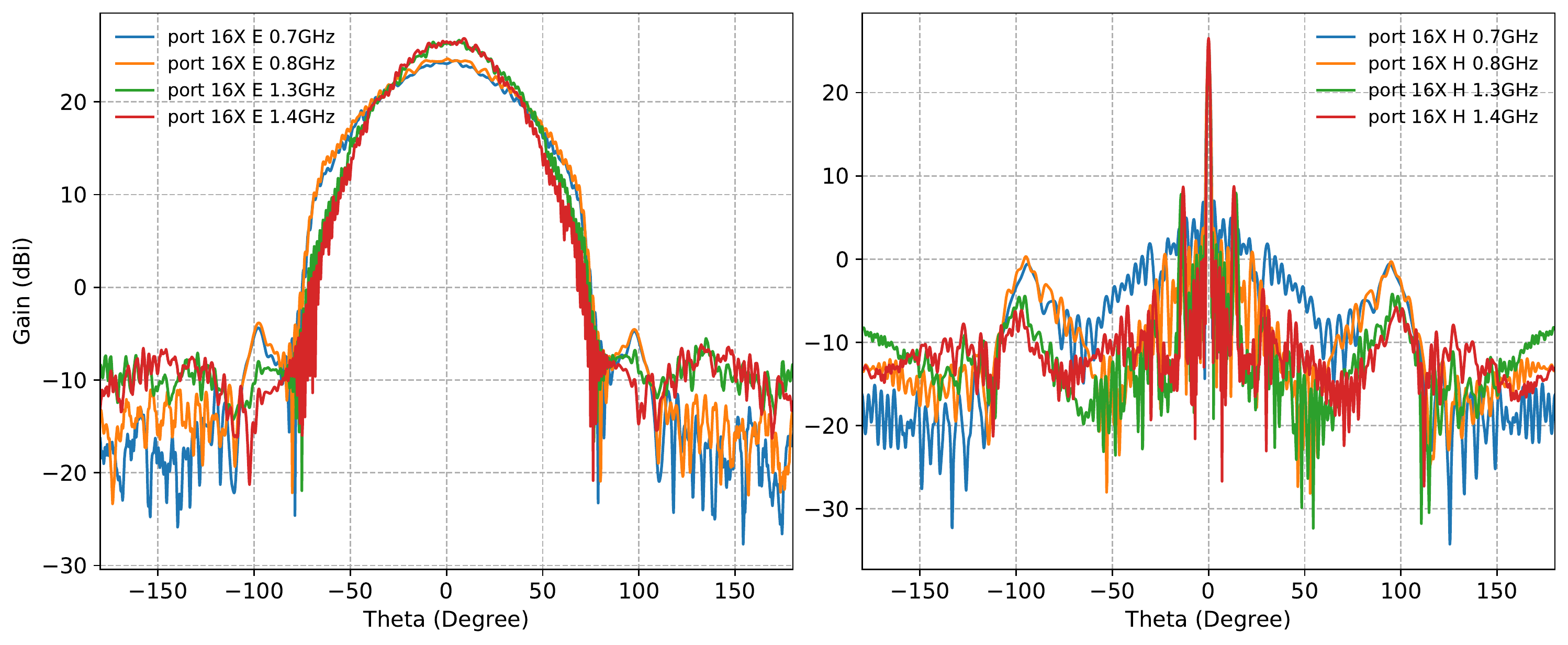}
    \caption{Simulated radiation pattern for the center feed Y-polarization (top panels) and X-polarization (bottom panels).  Left: E plane. Right: H plane.}
    \label{fig:31 feeds port1 E H plane radiation pattern}
\end{figure}
\begin{figure}[!htbp]
    \centering
	\includegraphics[width=0.9\columnwidth]{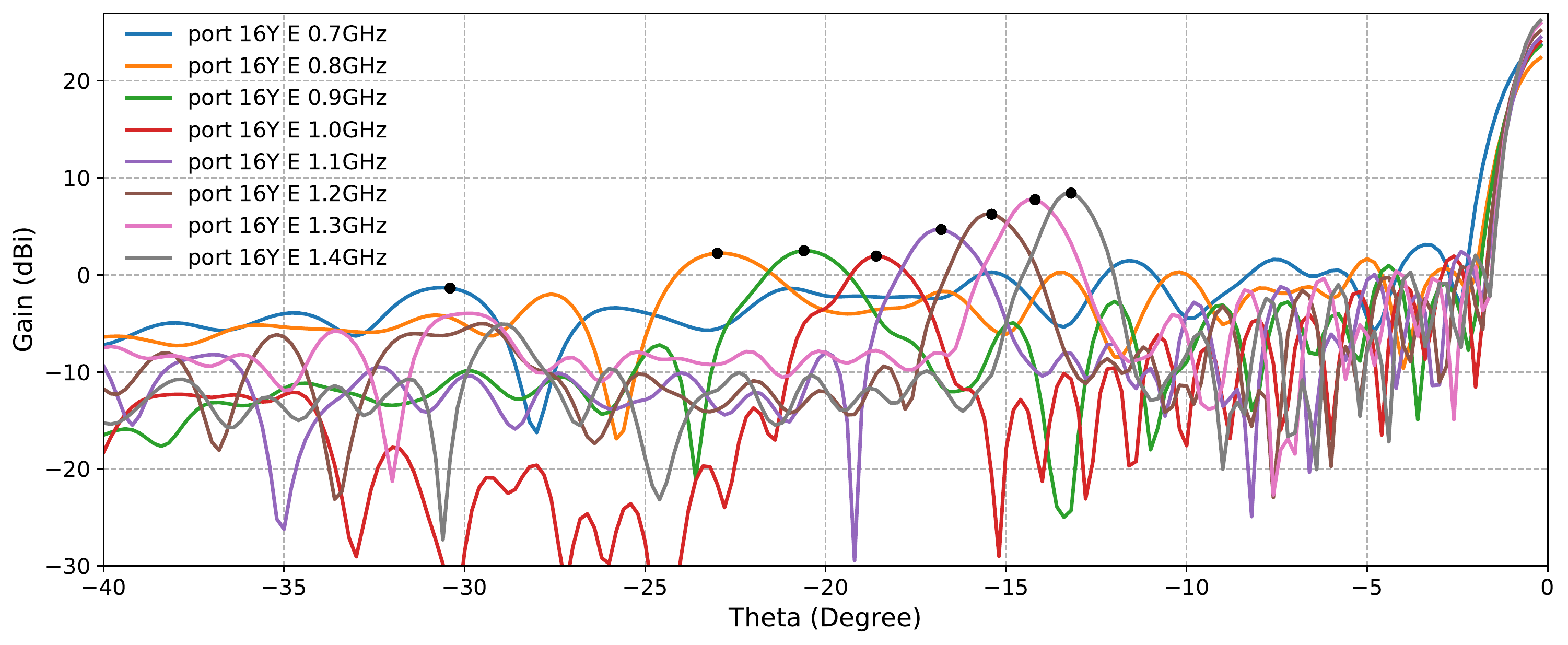}
    \caption{Simulated radiation pattern for Port 16Y and its maximum side lobe at 0.7 GHz - 1.4 GHz.}
    \label{fig:31 feeds port1 E sidelobe}
\end{figure}

The simulated radiation patterns of the center feed on the A cylinder are shown in Fig.~\ref{fig:31 feeds port1 E H plane radiation pattern} for a number of frequencies. The top and bottom panels show the Y and X polarization, and the left and right rows show the E plane and H plane respectively. $\theta=0^\circ$ is defined as the direction to the sky zenith. As expected for the cylinder reflector, the beam has a sharp peak along the East-West direction, while a broad one in the North-South direction, defined mostly by the beam of a single feed. 

Fig.~\ref{fig:31 feeds port1 E sidelobe} shows the pattern of the central feed E plane in the $\theta$ range of $-40^\circ \sim 0^\circ$. We can see that the position and the width of the peak varies significantly with frequency, and the peak amplitude for the highest peak also varies a lot. The highest side lobes are located at $10^\circ \sim 30^\circ$ away from the center of main lobe, some as high as 9 dBi. The majority of side lobes may be caused by additional reflections from the un-illuminated panels of the cylindrical reflector and by the non-uniformity of the aperture distribution. Nevertheless, these side lobes are still relatively low compared with the gain of the main beam. Comparing with the main lobe, the strongest side lobe is about -21.91 dB at 0.7 GHz, though it rises to about -16.98 dB at 1.4 GHz.

\begin{figure}[!htbp]
    \centering
	\includegraphics[width=0.85\columnwidth]{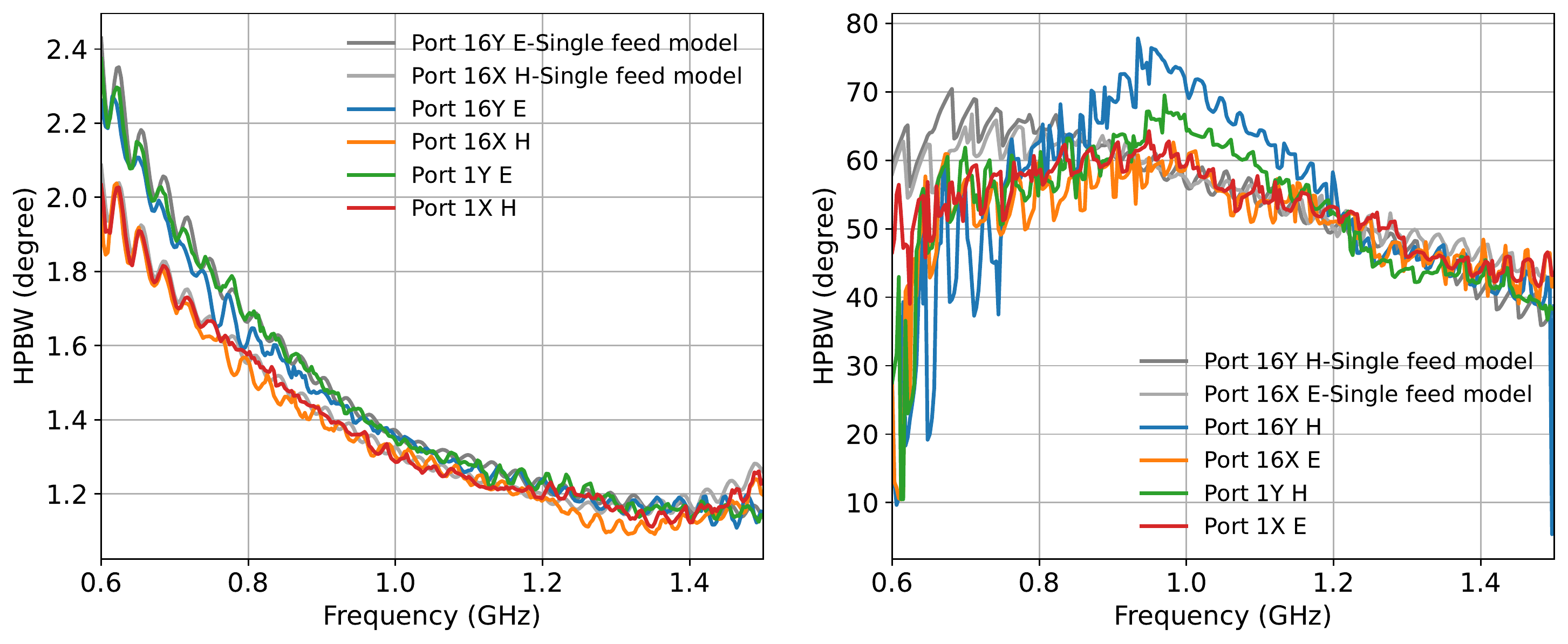}
	\includegraphics[width=0.5\columnwidth]{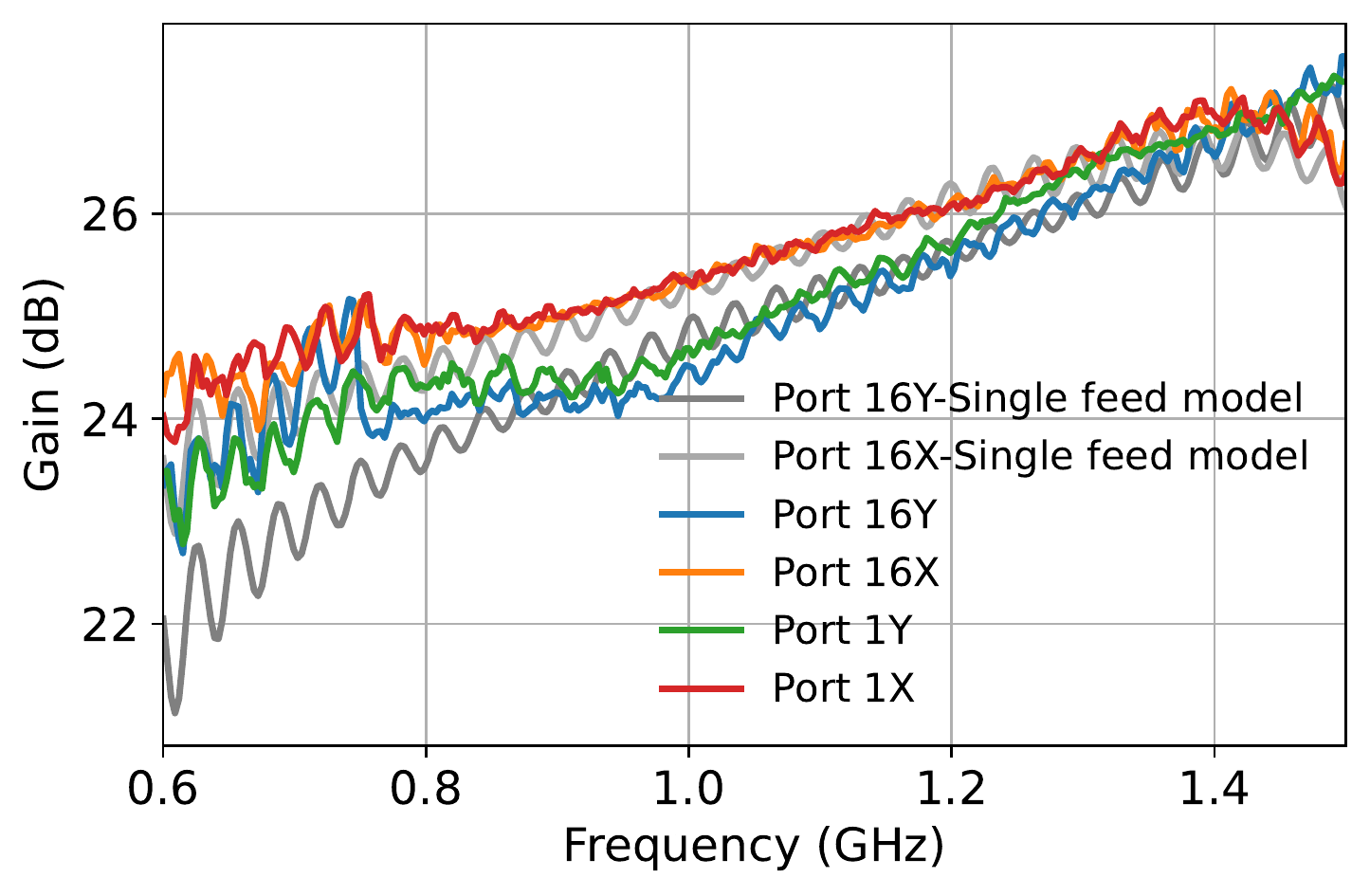}
    \caption{The HPBW (top panels) and peak gain (bottom panel) for the central feed and the feed at the edge with reflector. Top Left: Y direction. Top Right: X direction.}
    \label{fig:feed array with reflector HPBW E H}
\end{figure}

\begin{figure}[!htbp]
    \centering
	\includegraphics[width=0.7\columnwidth]{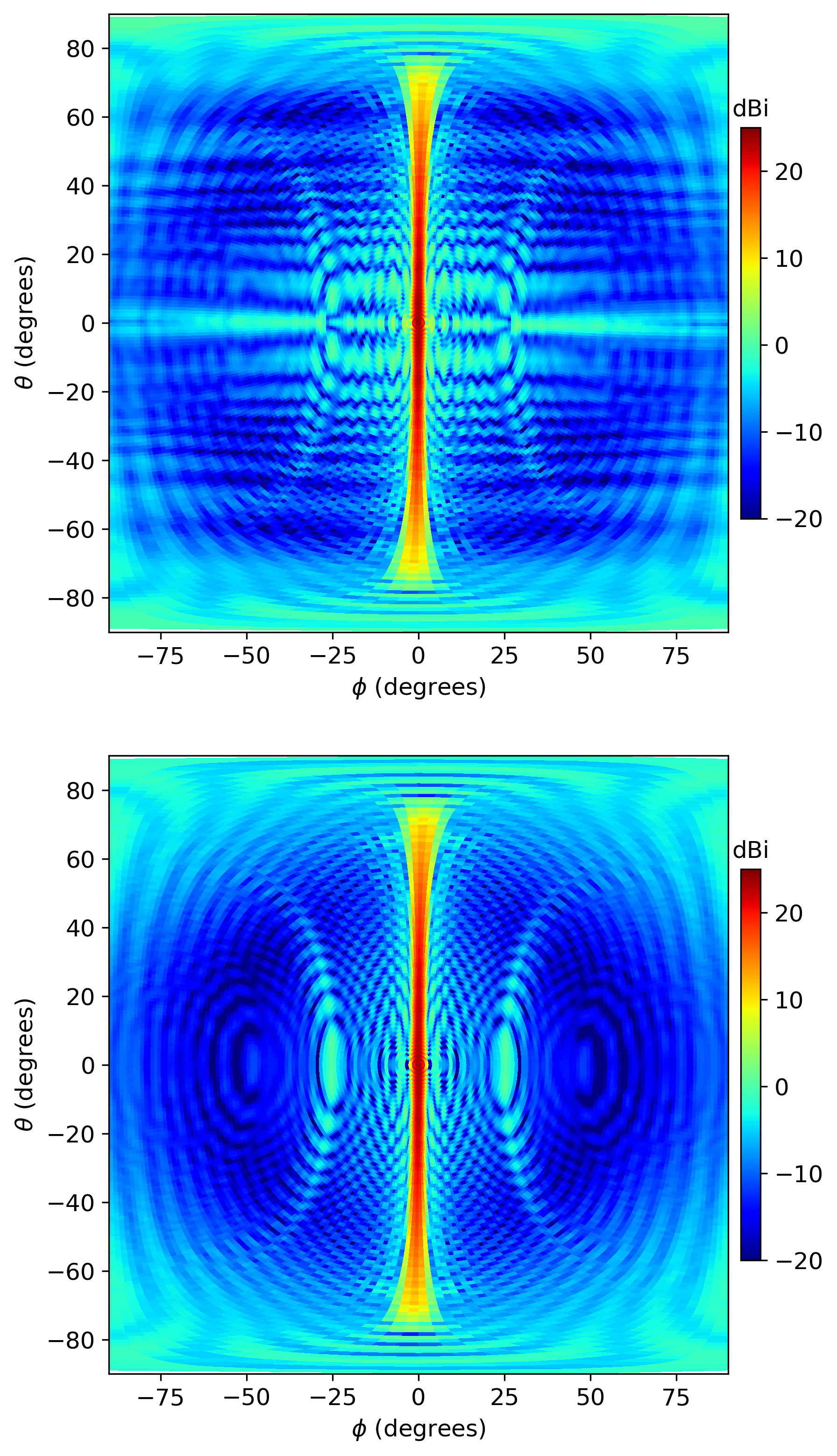}
    \caption{Simulated radiation pattern of the central feed Y-polarization at 0.75 GHz, here $\theta$ is the off-zenith angle, defined to be positive for north-inclined vectors and negative for south-inclined vectors, and the azimuth angle $\phi$ is measured with respect to the due north or south direction. Top panel: reflector with feed array model. Bottom panel: reflector with single feed model.}
    \label{fig:radiation pattern 2d comparision feed array and single feed}
\end{figure}

The HPBW of the feed at the center and at the edge are shown as a function of frequency in the top panels of Fig.~\ref{fig:feed array with reflector HPBW E H} for both polarizations.  It can be seen that the HPBW of the Y polarization decreases rapidly with frequency. The Top Right panel shows the HPBW in the X polarization. As a comparison, the HPBW of a single feed with a reflector model is also plotted; the single feed with reflector model is a good approximation. The peak gain of these two feeds are shown in the bottom panel of Fig.~\ref{fig:feed array with reflector HPBW E H}. As the frequency increases from 0.6 GHz to 1.5 GHz, the peak gains all increase.  The X polarization has higher gain than the Y polarization.

\begin{figure}[!htbp]
    \centering
	\includegraphics[width=\columnwidth]{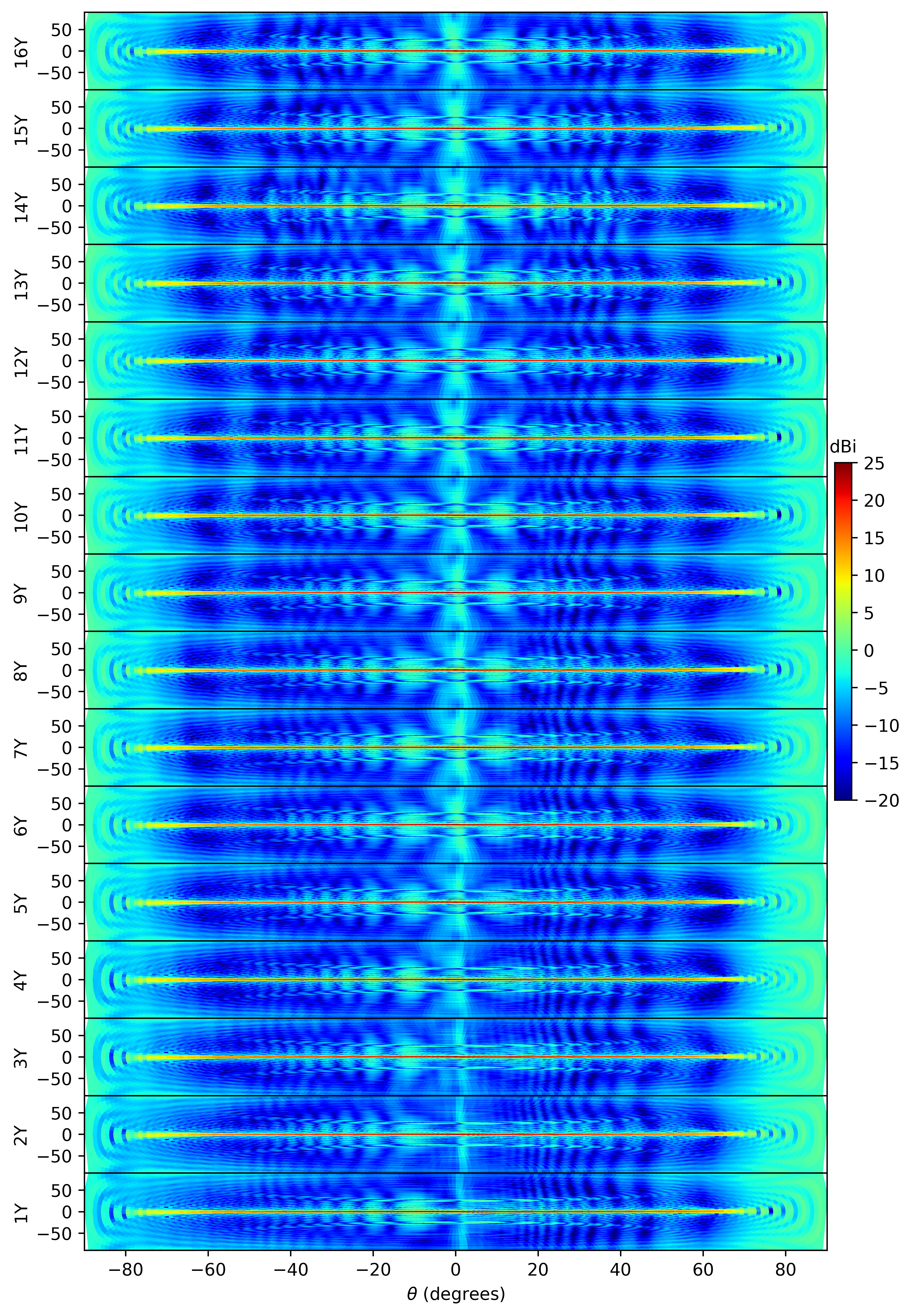}
    \caption{The 2D radiation patterns for the Y-polarization of NO.1 (edge) to No.16 (central) feeds at 0.75 GHz. For each feed, the horizontal and vertical axes show the $\theta$ and $\phi$ angle as defined above.}
    \label{fig:feed array radiation pattern contour odd}
\end{figure}

\begin{figure}[!htbp]
    \centering
	\includegraphics[width=\columnwidth]{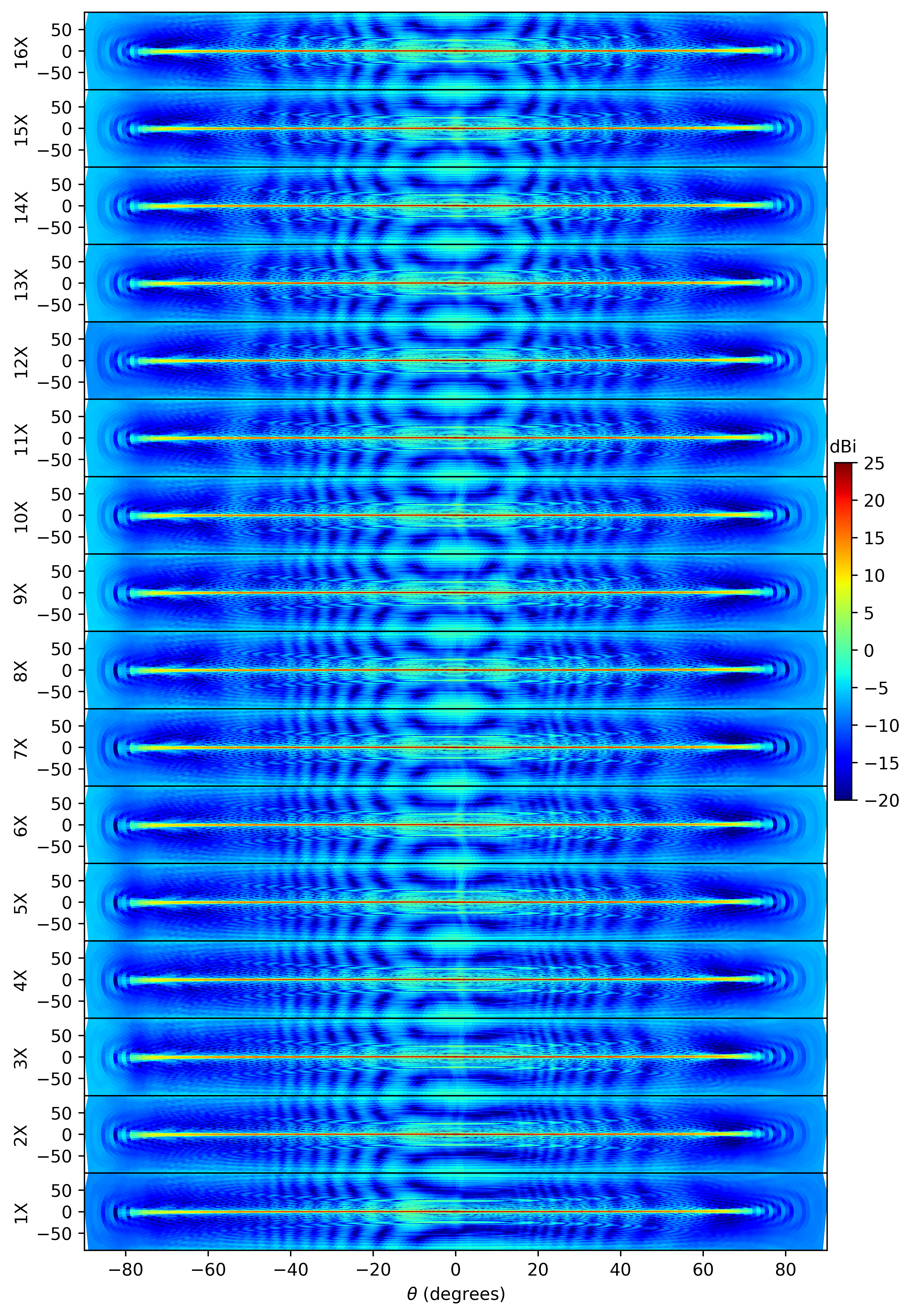}
    \caption{Same as the above figure, but for the X-polarization. 
    }
    \label{fig:feed array radiation pattern contour even}
\end{figure}

To fully appreciate the beam of the telescope, we show the 2D pattern in Fig.\ref{fig:radiation pattern 2d comparision feed array and single feed} for a frequency of 0.75 GHz. The central feed Y-polarization is used as illumination. Here we adopt a local telescope coordinate system used by CHIME \citep{deng2020antenna}, $\theta$ is the off-zenith angle, defined to be $0^\circ$ at zenith, positive(negative) for north(south) -inclined vectors, and $\phi$ is the azimuth angle measured with respect to either the due north (for $\theta>0^\circ$) or due south (for $\theta<0^\circ$) and defined in the range of $[-90^\circ,90^\circ]$. We also compare the radiation pattern of a single feed with reflector (bottom panel) and a feed in the feed array with the reflector (top panel). The main lobes for the two models are almost the same, but there are obvious differences in the side lobes. In the array case, the side lobes generally have greater amplitude, and are more extended, especially along the horizontal line with $\theta$ of $0^\circ$.

The occlusion between the feeds will cause distortion of the radiation pattern, we show the radiation pattern for all feeds at 0.75 GHz in Fig.~\ref{fig:feed array radiation pattern contour odd} for the Y-polarization, and Fig.~\ref{fig:feed array radiation pattern contour even} for the X-polarization. The horizontal and vertical axes show the $\theta$ angle and $\phi$ angle respectively, as defined above. We note that while there are general similarities in the overall pattern, there are subtle differences in the pattern for each feed. The difference in the far side lobes are more noticeable than the central peak. Also, the radiation pattern of the central feed (No.16) is highly symmetric to $\theta$ with respect to the zenith, but as we move away from the center, the pattern begin to show asymmetry with respect to $\theta$, especially for the side lobes. This is due to the asymmetric position of the feeds in the feed array.

The antenna efficiency $\eta$ can be written as the product of a series of terms \citep{thompson1999fundamentals, baars2007paraboloidal}

\begin{equation}
    \rm \eta=\eta_{r}\eta_{at}\eta_{so}\eta_{s}\eta_{si}\eta_{p}
    \label{eq:eta total}
\end{equation}
where $\eta_{r}$ is the radiation efficiency (typically very close to 1 as ohmic loss is small); $\eta_{at}$ is aperture tapering due to the under illumination of the reflector by the feed; $\eta_{so}$ is spill-over due to the beam pattern of the feed; $\eta_{s}$ is due to scattering off the structures blocking the surface; $\eta_{si}$ is surface irregularities, and $\eta_{p}$ is the surface roughness. The efficiency due to surface roughness $\eta_{p}$ is then given by
\begin{equation}
    \eta_{p}=\exp[-B(\frac{4\pi \epsilon}{\lambda})^2]
    \label{eq:eta p}
\end{equation}
where $B$ is a factor between 0 and 1 that depends on the radius of curvature, increasing as the radius of curvature decreases, and $\epsilon$ is the rms deviation of the reflector surface from a perfect paraboloid. Hence $\eta_{p}$ can be calculated by Eq.~(\ref{eq:eta p}), $\epsilon \sim 7.13$mm. $\eta_{p}$ will cause the total efficiency loss for 2.7\% (at 0.6 GHz), and 16\% (at 1.5 GHz). 

\begin{figure}[!htbp]
    \centering
	\includegraphics[width=0.6\columnwidth]{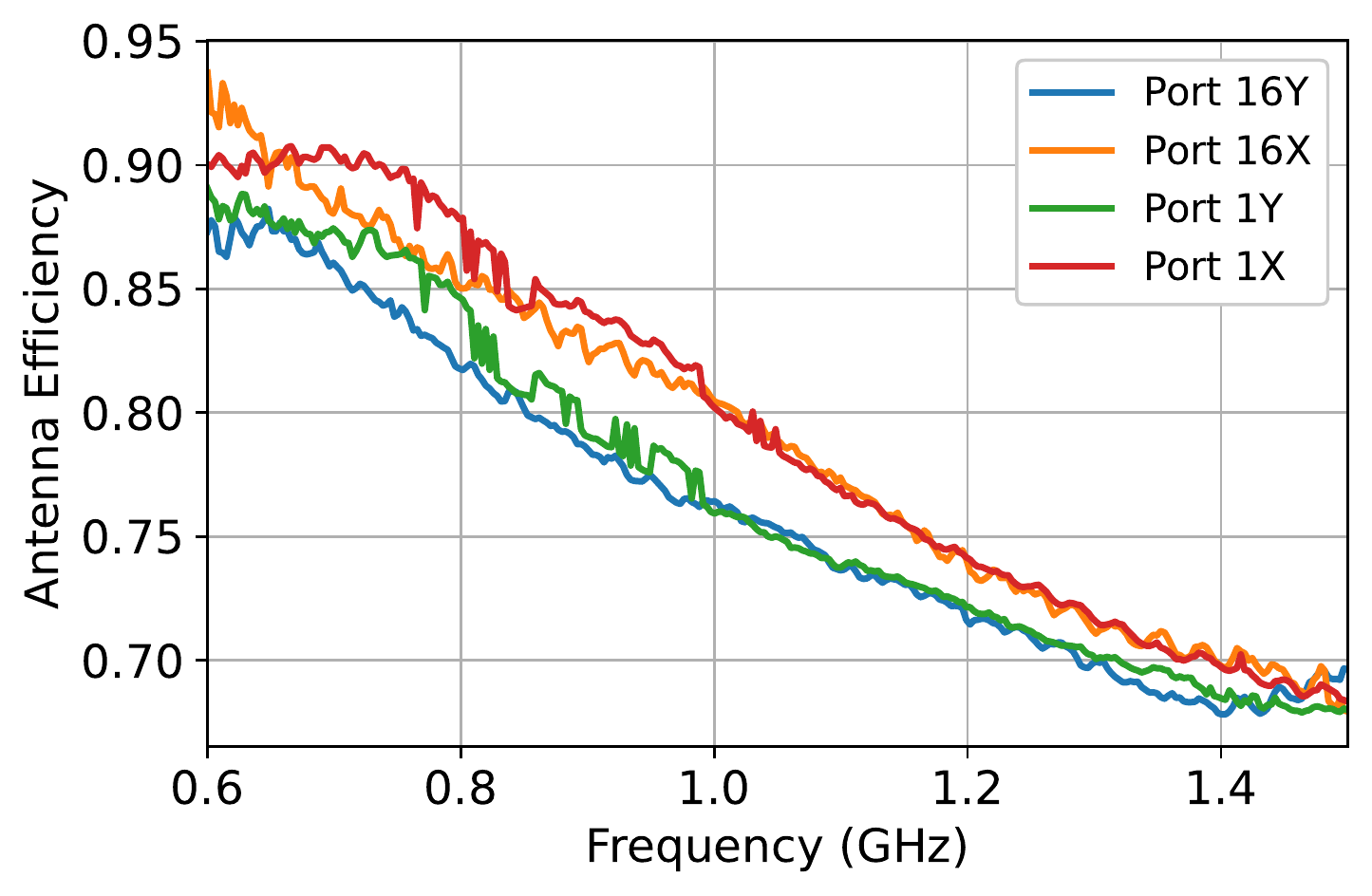}
    \caption{Antenna efficiency as a function of frequency for the feed at the center and the one at the edge.}
    \label{fig:radiation efficiency port1 2 31 32}
\end{figure}

The simulated total antenna efficiency of the feed at the center and at the edge are shown in Fig.~\ref{fig:radiation efficiency port1 2 31 32}. The efficiency varies from 90\% (at 0.6 GHz) to 68\% (at 1.5 GHz). Note that in the present simulation, we model the reflector as an ideal conductor, and with no aperture blockage, but in reality, some factors such as ohmic losses, surface irregularities, radiation leakage, and aperture occlusion could cause additional efficiency losses.

The frequency-dependent efficiency, and the oscillatory variation of the peak gain and beam shown in Fig.~\ref{fig:feed array with reflector HPBW E H} will all modulate the spectrum as seen from the end of receiver, thus contribute to the overall bandpass of the system. 
 
\section{Experimental measurement results}
\label{sec: observation results}

In this section, we compare our simulation with experiment data, which is available in the band of 0.7114 GHz - 0.7816 GHz. We will use the data 
collected in a period of 9 days starting on March 22nd, 2018.

\begin{figure}[!htbp]
    \centering
	\includegraphics[width=0.8\columnwidth]{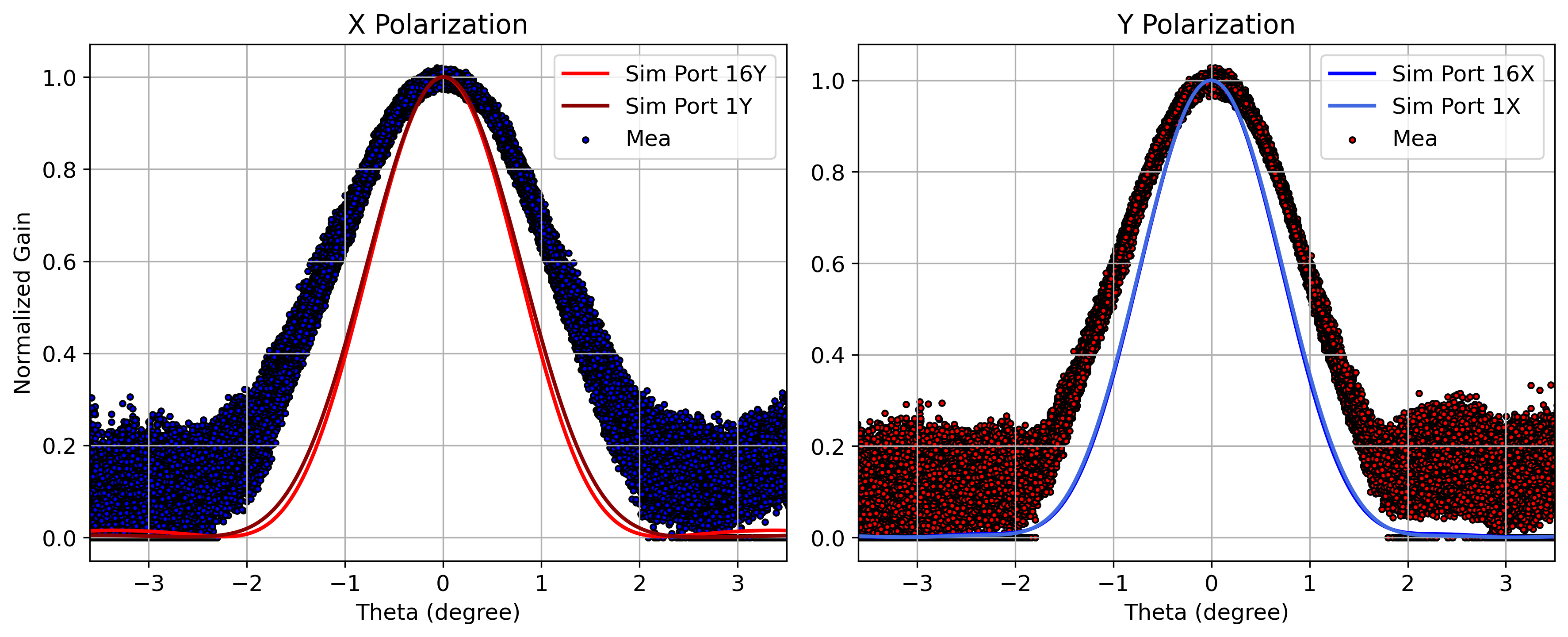}
    \caption{Measured and simulated beam pattern at 0.75 GHz. Left: X polarization. Right: Y polarization.}
    \label{fig:beam pattern}
\end{figure}

Since the Tianlai cylindrical telescope is static, to measure its beam profile, we observe bright celestial sources transiting through the field of view \citep{zuo2019eigenvector}. Cygnus A, which is located at ($19^h 59^m 28.3566^s, 40^\circ 44^\prime 02.096''$) in J2000 coordinates, is close to the zenith at its transit. Both polarizations of the measured and simulated beam patterns at 750 MHz are plotted as a function of angle in Fig.~\ref{fig:beam pattern}. The measured beam patterns are for all the 96 feeds, with nearly identical beam profile in the central part. In the side lobes where the measurement error is large, the profiles of different feeds vary a lot. If the measured beam pattern is fitted to a Gaussian function we can obtain the HPBW for the pattern. The HPBW is $2.54^\circ$ for Y polarization, and $2.1^\circ$ for X polarization. The measured beam width are wider than the simulation result, as shown in Fig.~\ref{fig:beam pattern}.

\begin{figure}[!htbp]
    \centering
	\includegraphics[width=0.6\columnwidth]{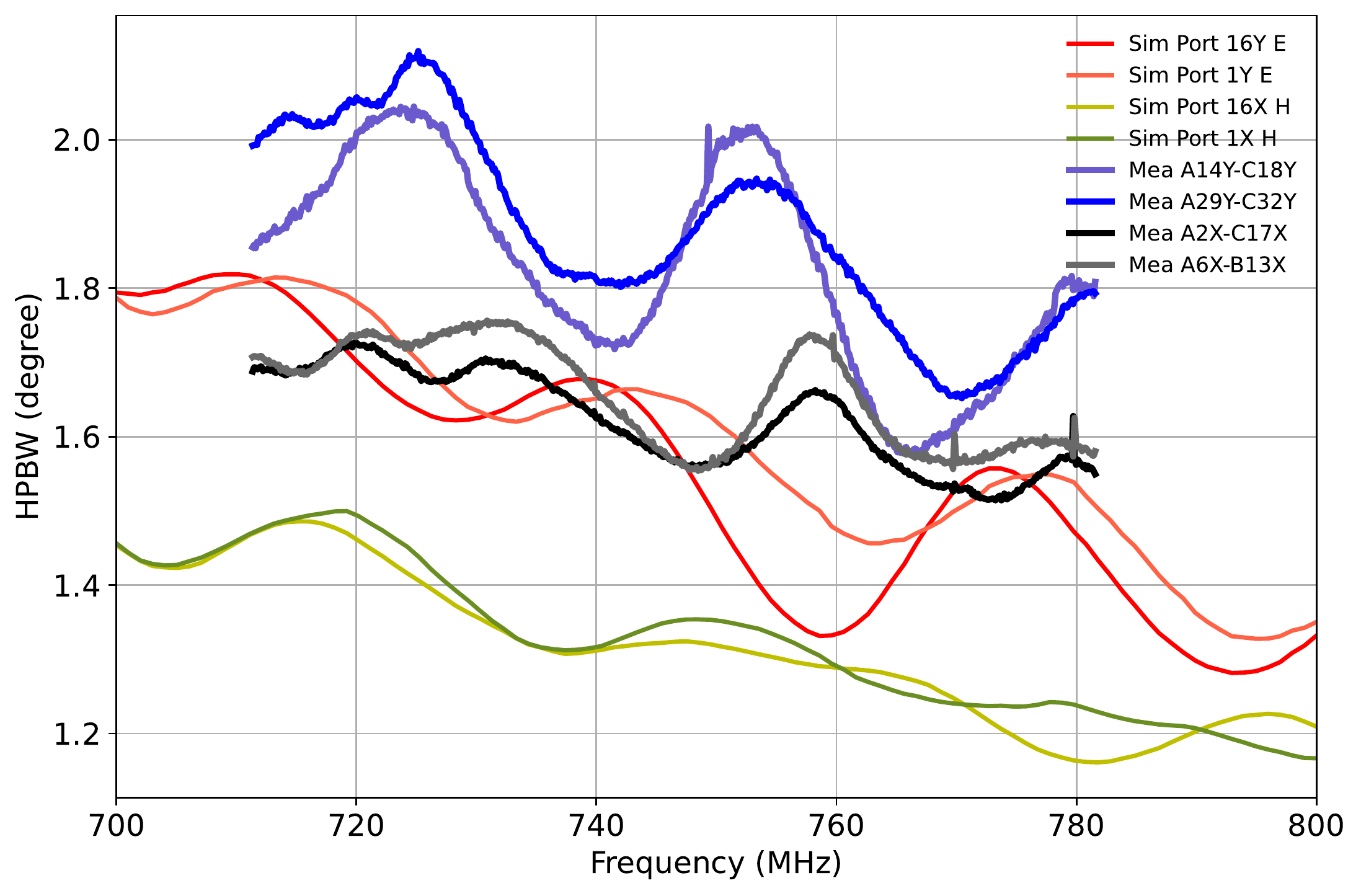}
    \caption{Measured and simulated HPBW versus frequency of 0.7 GHz -  0.8 GHz in Y direction.}
    \label{fig:HPBW E 0.7G to 0.8G obs and sim}
\end{figure}

We plot the simulated and measured HPBW in the Y direction of the cylindrical antenna as a function of frequency in Fig.~\ref{fig:HPBW E 0.7G to 0.8G obs and sim}. For the simulation, we choose the feed at the center and the feed at the edge. The simulation results show that the HPBW of the Y-polarized ports is $1.8^\circ$ at 0.7 GHz, and decreases with frequency, to $1.35^\circ$ at 0.8 GHz. HPBW for the X-polarized ports is $1.45^\circ$ at 0.7 GHz, and decreases with frequency, to $1.2^\circ$ at 0.8 GHz. For the measurement, we choose four pairs: A14Y-C18Y, A29Y-B32Y, A2X-C17X, and A6X-B13X. 
The HBPW is then measured repeatedly in the East-West direction by observing the transit of Cygnus A. For each transit a Gaussian shape is fitted to the magnitude of the visibility as a function of time. Comparing the simulated and measured results, the HPBW in both the X- and Y- polarization decrease with frequency, and the HPBW for the Y-polarization is $~0.3^\circ$ wider than the HPBW for the X-polarization. The HPBW curves show oscillatory features versus frequency in both the simulated and measured results.

\begin{figure}[!htbp]
    \centering
	\includegraphics[width=0.8\columnwidth]{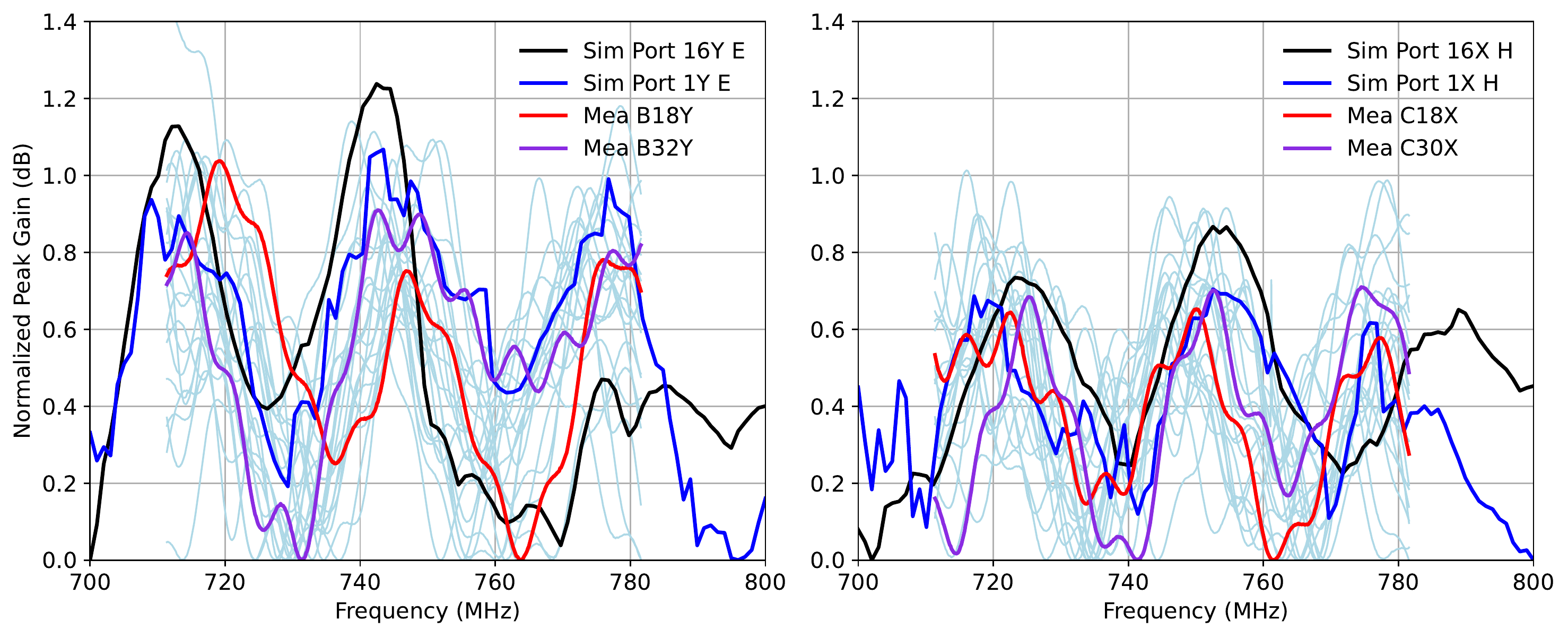}
    \caption{Measured and simulated peak gain verse frequency of 0.7 GHz - 0.8 GHz. For measured results, several channels of autocorrelation data is shown as light blue lines, four typical channels B18Y, B32Y, C18X, C30X are highlighted. Left: Y-polarized ports. Right: X-polarized ports.}
    \label{fig:feed array peak gain 700 to 800 comparision}
\end{figure}

The peak gain from both simulation and measurement are plotted as a function of frequency in Fig.~\ref{fig:feed array peak gain 700 to 800 comparision}. For simulation, we still use these typical four ports. For measurement, we choose several channels of autocorrelation data for both X-polarization and Y-polarization, and four typical channels are highlighted: B18Y is close to the center in reflector B, B32Y is in the edge in reflector B; C18X is close to the center in reflector C, and C30X is close to the edge in reflector C. We note that despite some difference in the absolute magnitude, the simulation and measurement results are similar in shape. The difference between the maximum and minimum values is $~1$ dB for Y polarization, and $~0.8$ dB for X polarization. Both the simulated and measured response have three peaks in the observation band with an interval of 31 MHz due to the reflection between the radiation blade of the feed and the reflector. For other similar channel pairs the results are similar.

\begin{figure}[!htbp]
    \centering
	\includegraphics[width=\columnwidth]{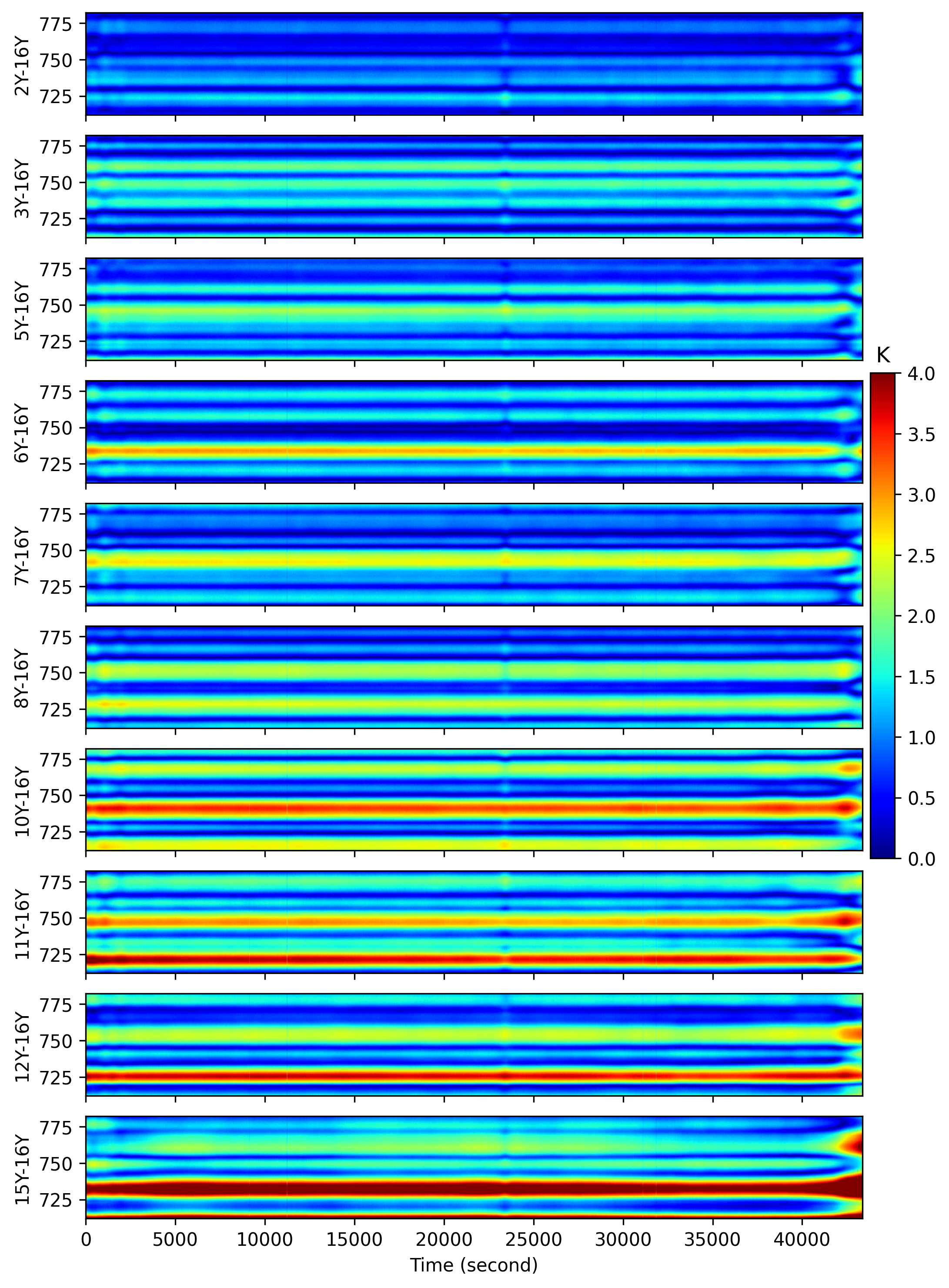}
    \caption{Visibilities as a function of time and frequency for several baselines. The data are rebinned to 488 kHz and 20 s resolution.}
    \label{fig:observation cross talk}
\end{figure}

Fig.~\ref{fig:observation cross talk} shows the waterfall plot of the visibility (cross correlation) data corresponding to multiple baselines on cylinder A. The scale is determined by comparing with the observation during the transit of calibrator source Cygnus A. We can see bright horizontal bands, which show noise which is persistent at all times. In an interferometer, purely random noise at the two input elements is uncorrelated and averages to zero. This persistent obviously does not; we shall call this the ``correlated noise". The mutual coupling between the different ports could contribute to such correlated noise, which is basically steady with both time and frequency, though some may also come from other sources, such as the ground pick up. The noise induced by the cross-couplings is coherent and persistent in the visibility data, and not easily reduced by long integration times. We see in the data that this noise is stronger for the shorter baselines, e.g. 15Y-16Y, the coupled noise temperature reaches $\sim 4$ K at some frequency points, and decreases with increasing baseline length, while for a baseline of 2Y-16Y, the coupled noise temperature is $\sim 1$ K. This noise also has a complicated frequency-dependence, so it is an important source of noise to 21 cm intensity mapping observations.

\begin{figure}[!htbp]
    \centering
	\includegraphics[width=\columnwidth]{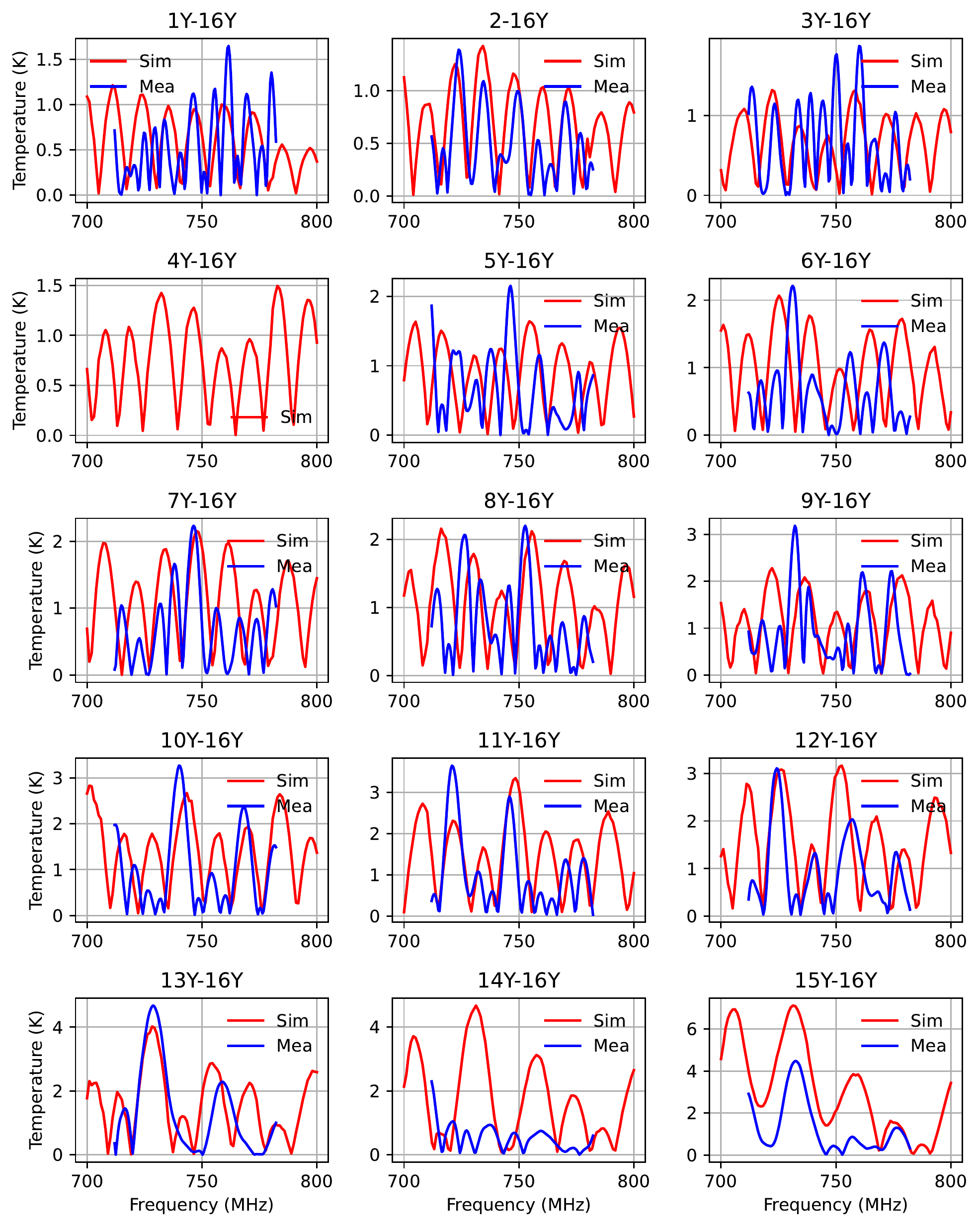}
    \caption{Visibilities (blue) and simulated mutual coupling noise temperature (red) as a function of frequency for different baselines (Port 4Y had malfunctioned at the time).}
    \label{fig:observation and measurement cross talk}
\end{figure}

The mutual coupling effects between feed elements can be characterized by transmission coefficients. The visibility is
\begin{eqnarray}
V_{ij} &=& \langle  \tilde{U}_i^* \tilde{U}_j \rangle  \nonumber \\
& = &\langle (U_i + S_{ij} U_j)^*  (U_j + S_{ji} U_i) \rangle \nonumber\\
& = & (1+{S_{ij}}^* S_{ji}) \langle U_i^* U_j \rangle + {S_{ij}}^* \langle |U_j|^2 \rangle + S_{ji}\langle |U_i|^2 \rangle 
\end{eqnarray}
where we used an perturbative approach, so  $\tilde{U_i}$ denotes the output voltage, while $U_i$ denotes the quantity associated with the the no-coupling case. The transmission coefficients $S_{ij} \ll 1$, so we can neglect the $S_{ij} S_{ji}$ term. Here we also neglect the slight difference in $U_i$, $U_j$ for different feeds $i,j$. The Tianlai feed units before the low noise amplifier (LNA) can be regarded as a passive, symmetric system, so that $S_{ij}=S_{ji}$, and $S_{ij}^*+S_{ji} = 2 {\rm Re}(S_{ij})$.  Then
\begin{equation}
\begin{aligned}
    V_{ij} = \langle U_i^* U_j \rangle + 2 {\rm Re}[S_{ij}]  \langle |U_i|^2 \rangle .
    \label{eq:equivalent system temperature 2}
\end{aligned}
\end{equation}
where the first term would be obtained without cross-coupling, while the second term is a kind of correlated noise arising from the cross-couplings, which is related to the 
total voltages in the feed. This is characterized by an antenna temperature,  $\langle |U_i|^2 \rangle=C T_{\rm ant}$, where $C$ is a coefficient.

To estimate this antenna temperature, we note that in the observed system temperature, there are contributions from the antenna and contributions from the receiver, so that 
\begin{equation}
T_{\rm sys} = T_{\rm ant} + T_{\rm rec}
\end{equation}
where $T_{\rm rec}$ is the noise temperature of the receiver.  The measured average system temperature for the A, B, C cylinders are 99.7 K, 85.8 K and 82.0 K respectively \citep{li2020tianlai}, though of course it also varies with time and different feeds. The receiver noise is dominated by that of the LNA; the measured mean noise figure (NF) of the LNA is 0.65 dB, with a variation of about 0.1 dB in the  700 MHz - 800 MHz band. The corresponding receiver noise temperature is given by 
\begin{equation}
T_{\rm rec} = T_{\rm amb} (10^{0.1 \rm NF}-1)  
\end{equation}
which is about 48 K for an ambient temperature of $T_{\rm amb}= 300 \K$. If we take a system temperature of 90 K, the antenna temperature is estimated to be 42 K. We can then use the antenna temperature and the coupling coefficients to estimate the magnitude of the correlated noise. 

In Fig.~\ref{fig:observation and measurement cross talk} we plot the simulated and measured mutual coupling noise temperature for several baselines as a function of frequency. For measurement, we use the mean of 11 hours of calibrated visibility (cross correlation) data, which is rebinned to 488 kHz and 20 s resolution. 
There is a generally good agreement between simulation and measurement results,  both in the magnitude of the noise temperature, and in the general frequency structure. There are still differences between the simulation and measurement results, but this is expected, and part of this may be due to the differences between the simulation model and the actual antenna structure, and the ground pick up contribution which may also contribute to the correlated noise but is not considered here.

\section{Conclusions}
\label{sec: conclusions}

The 21 cm observation is a very promising cosmological probe, but detecting the weak cosmological 21 cm signal out of the huge foreground is also exceedingly hard, which requires unprecedented precision in calibration. Observation results from the Tianlai cylinder pathfinder show that there are oscillatory features in the system bandpass response that are associated with standing wave resonances, and mutual coupling effects between different feeds. These all make the data analysis more complicated.  In this paper, we investigated the Tianlai cylindrical telescope performance with electromagnetic simulations, and compared the results with measurement data.

Electromagnetic simulation of the Tianlai array is itself very challenging, due to the large and complicated structure, and the vast scales which range from the sub-wavelength geometry of the feed, to the much larger reflector structure. The simulation requires much computing resource and is very time-consuming. We divide the whole simulation into three parts, and use different kinds of methods according to the model size: FEM for the single feed model, FE-BI for the feed array and PO for the feed array with the cylindrical reflector. The optimized feed simulation model is demonstrated to be precise by comparing the simulation results with the measurements in a microwave anechoic chamber. We then simulate a linear array of feeds, and characterize their mutual coupling effects using the S-parameters. Finally, we put in the reflector. The coupling is much enhanced with the cylindrical reflector; even between the central feed and the feed at the edge it is -40 dB, but there is still a trend of decrease with increasing distance between the feeds. This coupling can be a major contribution to the persistent correlated noise seen in the interferometer visibility data. 

We obtain radiation patterns for the different feeds on the cylinder. The central feed has a symmetric pattern, while for the feeds on the ends, the side lobes show asymmetry although the main lobe is still largely symmetric. The beam width decreases with frequency, and the beams are wider for the Y-polarization than for the X-polarization. The beam width derived from observational data seems to be wider at the 10\% level than given by the simulation. Despite this,  the simulation does reveal some features seen in the observational data: the HPBW and peak gain show oscillatory features along frequency, and both the simulation and measurement show three peaks with an interval of $\sim$ 31 MHz, which is associated with the signal reflection between the feed and the cylindrical reflector. We also estimated the correlated noise, which is determined by the coupling coefficients and antenna temperature. We find the estimated correlated noise in visibility has a magnitude comparable with the actual data, and the frequency structure also agrees with the data qualitatively. 

There are a number of simplifications and limitations in the present simulation. We have limited our simulation to a single cylinder, though the three cylinders are next to each other, and there is also the array of 16 dish antennas nearby, which may affect the result. The large cylindrical antennas have supporting struts and a frame for anchoring the feed units, which have been omitted in this model.  We have treated the feed and reflector as ideal, neglecting possible manufacturing and installation errors. Taking into account these issues would improve the realism of the simulation, but would require much more computing resource. 

Despite these limitations, the present simulation reproduces many features seen in the observational data, at least at the qualitative level, which helps us to understand the origin of these features. The complicated frequency response and couplings between the different elements need to be taken into account and mitigated for high precision 21cm observations.

\section*{Acknowledgements}
We thank Jeff Peterson and John Marriner for helpful discussions. 
The Tianlai arrays are operated with the support of the National Astronomical Observatory of China (NAOC) Astronomical Technology Center. This work is supported by the Ministry of the Science Technology  MoST-BRICS Flagship Project 2018YFE0120800, National SKA Program of China No. 2020SKA0110401, the National Key R\&D Program 2017YFA0402603; the National Natural Science Foundation of China (NSFC) grant 11973047, 11633004, U1631118, the Chinese Academy of Sciences (CAS) Strategic Priority Research Program XDA15020200, the CAS Frontier Science Key Project QYZDJ-SSW-SLH017,
and the CAS Inter-disciplinary Innovation Team (JCTD-2019-05), the CAS Key Instruments project ZDKYYQ20200008, and the Hebei Key Laboratory of Radio Astronomy Technology (HKLRAT).


\bibliographystyle{raa}
\bibliography{references}  

\begin{thebibliography}{43}
\providecommand\natexlab[1]{#1}
\providecommand\JournalTitle[1]{#1}

\bibitem[Abdalla {et~al.}(2021)]{abdalla2021bingo}
Abdalla, E., Ferreira, E.~G., Landim, R.~G., {et~al.} 2021, arXiv preprint
  arXiv:2107.01633

\bibitem[Amiri {et~al.}(2022{\natexlab{a}})]{amiri2022detection}
Amiri, M., Bandura, K., Chen, T., {et~al.} 2022{\natexlab{a}}, arXiv preprint
  arXiv:2202.01242

\bibitem[Amiri {et~al.}(2022{\natexlab{b}})]{amiri2022overview}
Amiri, M., Bandura, K., Boskovic, A., {et~al.} 2022{\natexlab{b}}, arXiv
  preprint arXiv:2201.07869

\bibitem[Ansari {et~al.}(2008)]{ansari2008reconstruction}
Ansari, R., Goff, J.-M.~L., Magneville, C., {et~al.} 2008, arXiv preprint
  arXiv:0807.3614

\bibitem[ANSYS(2021)]{HFSS}
ANSYS, I. 2021, HFSS

\bibitem[Baars(2007)]{baars2007paraboloidal}
Baars, J.~W. 2007, The paraboloidal reflector antenna in radio astronomy and
  communication, Vol. 348 (Springer)

\bibitem[Bacon {et~al.}(2020)]{SKA:2018ckk}
Bacon, D.~J., {et~al.} 2020, Publ. Astron. Soc. Austral., 37, e007

\bibitem[Bandura {et~al.}(2019)]{bandura2019packed}
Bandura, K., Castorina, E., Connor, L., {et~al.} 2019, Packed Ultra-wideband
  Mapping Array (PUMA): A Radio Telescope for Cosmology and Transients,
  arXiv:1907.12559

\bibitem[Chang {et~al.}(2008)]{chang2008baryon}
Chang, T.-C., Pen, U.-L., Peterson, J.~B., \& McDonald, P. 2008, Physical
  Review Letters, 100, 091303

\bibitem[Chen(2012)]{chen2012tianlai}
Chen, X. 2012, in International Journal of Modern Physics: Conference Series,
  Vol.~12, World Scientific, 256

\bibitem[Chen(2015)]{chen2015AAPPS}
Chen, X. 2015, Asso. Asia Pac. Phys. Soc. Bull., 25

\bibitem[Chen {et~al.}(2015)]{chen2015tianlai}
Chen, X., Wu, F., Shi, H., Zhang, J., \& Wang, Y. 2015, IAU General Assembly,
  22, 2252187

\bibitem[Cianciara {et~al.}(2017)]{cianciara2017simulation}
Cianciara, A.~J., Anderson, C.~J., Chen, X., {et~al.} 2017, Journal of
  Astronomical Instrumentation, 6, 1750003

\bibitem[Crichton(2018)]{crichton2018hydrogen}
Crichton, D.~T. 2018, in 2018 2nd URSI Atlantic Radio Science Meeting
  (AT-RASC), IEEE, 1

\bibitem[Das {et~al.}(2018)]{Das2018}
Das, S., Anderson, C.~J., Ansari, R., {et~al.} 2018, in Millimeter,
  Submillimeter, and Far-Infrared Detectors and Instrumentation for Astronomy
  IX, Vol. 10708, International Society for Optics and Photonics, 1070836

\bibitem[DeBoer {et~al.}(2017)]{HERA2017}
DeBoer, D.~R., Parsons, A.~R., Aguirre, J.~E., {et~al.} 2017, Publications of
  the Astronomical Society of the Pacific, 129, 045001

\bibitem[Deng(2020)]{deng2020antenna}
Deng, M. 2020, Antenna array design, beam calibration of the CHIME to measure
  the late-time cosmic acceleration and mapping of the north celestial cap, PhD
  thesis, University of British Columbia

\bibitem[Edgar(2011)]{edgar2011hfss}
Edgar, D. 2011

\bibitem[Hotan {et~al.}(2021)]{hotan2021australian}
Hotan, A., Bunton, J., Chippendale, A., {et~al.} 2021, Publications of the
  Astronomical Society of Australia, 38

\bibitem[Hu {et~al.}(2020)]{Hu:2019okh}
Hu, W., Wang, X., Wu, F., {et~al.} 2020, Mon. Not. Roy. Astron. Soc., 493, 5854

\bibitem[Jin(2015)]{jin2015finite}
Jin, J.-M. 2015, The finite element method in electromagnetics (John Wiley \&
  Sons)

\bibitem[Johnston {et~al.}(2008)]{johnston2008science}
Johnston, S., Taylor, R., Bailes, M., {et~al.} 2008, Experimental astronomy,
  22, 151

\bibitem[Jonas \& Team(2016)]{jonas2016meerkat}
Jonas, J., \& Team, M. 2016, MeerKAT Science: On the Pathway to the SKA, 1

\bibitem[Kern {et~al.}(2019)]{Kern_2019a}
Kern, N.~S., Parsons, A.~R., Dillon, J.~S., {et~al.} 2019, The Astrophysical
  Journal, 884, 105

\bibitem[{Kern} {et~al.}(2020)]{Kern_2019b}
{Kern}, N.~S., {Parsons}, A.~R., {Dillon}, J.~S., {et~al.} 2020, \apj, 888, 70

\bibitem[{Koopmans} {et~al.}(2015)]{Koopmans2015}
{Koopmans}, L., {Pritchard}, J., {Mellema}, G., {et~al.} 2015, Advancing
  Astrophysics with the Square Kilometre Array (AASKA14), 1

\bibitem[Li {et~al.}(2021)]{li2021reflections}
Li, J.-X., Wu, F.-Q., Sun, S.-J., {et~al.} 2021, Research in Astronomy and
  Astrophysics, 21, 059

\bibitem[Li {et~al.}(2020)]{li2020tianlai}
Li, J., Zuo, S., Wu, F., {et~al.} 2020, SCIENCE CHINA Physics, Mechanics \&
  Astronomy, 63, 1

\bibitem[Liu {et~al.}(2014)]{liu2014design}
Liu, T., Wu, F., Chen, Z., {et~al.} 2014, in 2014 XXXIth URSI General Assembly
  and Scientific Symposium (URSI GASS), IEEE, 1

\bibitem[Newburgh {et~al.}(2016)]{newburgh2016hirax}
Newburgh, L., Bandura, K., Bucher, M., {et~al.} 2016, in Ground-based and
  Airborne Telescopes VI, Vol. 9906, International Society for Optics and
  Photonics, 99065X

\bibitem[O'Connor {et~al.}(2020)]{oconnor2020baryon}
O'Connor, P., Slosar, A., Harris, M., {et~al.} 2020, The Baryon Mapping
  Experiment (BMX), a 21cm intensity mapping pathfinder, arXiv:2011.08695

\bibitem[Parsons {et~al.}(2010)]{PAPER2010}
Parsons, A.~R., Backer, D.~C., Foster, G.~S., {et~al.} 2010, The Astronomical
  Journal, 139, 1468

\bibitem[Seo {et~al.}(2010)]{seo2010ground}
Seo, H.-J., Dodelson, S., Marriner, J., {et~al.} 2010, The Astrophysical
  Journal, 721, 164

\bibitem[Silvestro(2010)]{silvestro2010hybrid}
Silvestro, J. 2010, Ansys, Inc.[online]. Available: https://support. ansys. com

\bibitem[Thompson(1999)]{thompson1999fundamentals}
Thompson, A.~R. 1999, in Synthesis Imaging in Radio Astronomy II, Vol. 180, 11

\bibitem[{Tingay} {et~al.}(2013)]{MWA2013}
{Tingay}, S.~J., {Goeke}, R., {Bowman}, J.~D., {et~al.} 2013, \pasa, 30, e007

\bibitem[{van Haarlem} {et~al.}(2013)]{LOFAR2013}
{van Haarlem}, M.~P., {Wise}, M.~W., {Gunst}, A.~W., {et~al.} 2013, \aap, 556,
  A2

\bibitem[Vanderlinde {et~al.}(2014)]{vanderlinde2014canadian}
Vanderlinde, K., Collaboration, C., {et~al.} 2014, Exascale Radio Astronomy, 2,
  10102

\bibitem[Wu {et~al.}(2021)]{wu2021tianlai}
Wu, F., Li, J., Zuo, S., {et~al.} 2021, Monthly Notices of the Royal
  Astronomical Society, 506, 3455

\bibitem[{Xu} {et~al.}(2015)]{Xu2015}
{Xu}, Y., {Wang}, X., \& {Chen}, X. 2015, \apj, 798, 40

\bibitem[{Zhang} {et~al.}(2016)]{2016RAA....16..158Z}
{Zhang}, J., {Zuo}, S.-F., {Ansari}, R., {et~al.} 2016, \raa, 16, 158

\bibitem[Zhao \& Petersson(2018)]{zhao2018overview}
Zhao, K., \& Petersson, L.~R. 2018, in 2018 IEEE International Symposium on
  Antennas and Propagation \& USNC/URSI National Radio Science Meeting, IEEE,
  411

\bibitem[Zuo {et~al.}(2019)]{zuo2019eigenvector}
Zuo, S., Pen, U.-L., Wu, F., {et~al.} 2019, The Astronomical Journal, 157, 34

\end{thebibliography}

\end{document}